# Beyond Bandlimited Sampling: Nonlinearities, Smoothness and Sparsity

Y. C. Eldar and T. Michaeli

Digital applications have developed rapidly over the last few decades. Since many sources of information are of analog or continuous-time nature, discrete-time signal processing (DSP) inherently relies on sampling a continuous-time signal to obtain a discrete-time representation. Consequently, sampling theories lie at the heart of signal processing devices and communication systems. Examples include sampling rate conversion for software radio [1] and between audio formats [2], biomedical imaging [3], lens distortion correction and the formation of image mosaics [4], and super-resolution of image sequences [5].

To accommodate high operating rates while retaining low computational cost, efficient analog-to-digital (ADC) and digital-to-analog (DAC) converters must be developed. Many of the limitations encountered in current converters is due to a traditional assumption that the sampling stage needs to acquire the data at the Shannon-Nyquist rate, corresponding to twice the signal bandwidth [6], [7], [8]. To avoid aliasing, a sharp low-pass filter (LPF) must be implemented prior to sampling. The reconstructed signal is also a bandlimited function, generated by integer shifts of the sinc interpolation kernel.

A major drawback of this paradigm is that many natural signals are better represented in alternative bases other than the Fourier basis [9], [10], [11], or possess further structure in the Fourier domain. In addition, ideal pointwise sampling, as assumed by the Shannon theorem, cannot be implemented. More practical ADCs introduce a distortion which should be accounted for in the reconstruction process. Finally, implementing the infinite sinc interpolating kernel is difficult, since it has slow decay. In practice, much simpler kernels are used such as linear interpolation. Therefore there is a need to develop a general sampling theory that will accommodate an extended class of signals beyond bandlimited functions, and will account for the nonideal nature of the sampling and reconstruction processes.

Sampling theory has benefited from a surge of research in recent years, due in part to the intense research in wavelet theory and the connections made between the two fields. In this survey we present several extensions of the Shannon theorem, that have been developed primarily in the past two decades, which treat a wide class of input

Department of Electrical Engineering, Technion—Israel Institute of Technology, Haifa 32000, Israel. E-mail: {yonina@ee,tomermic@tx}.technion.ac.il. Fax: +972-4-8295757. Tel: +972-4-8293256.

This work was supported in part by the Israel Science Foundation under Grant no. 1081/07 and by the European Commission in the framework of the FP7 Network of Excellence in Wireless COMmunications NEWCOM++ (contract no. 216715).

signals as well as nonideal sampling and nonlinear distortions. This framework is based on viewing sampling in a broader sense of projection onto appropriate subspaces, and then choosing the subspaces to yield interesting new possibilities. For example, our results can be used to uniformly sample non-bandlimited signals, and to perfectly compensate for nonlinear effects.

Our focus here is on shift-invariant (SI) settings in which both sampling and reconstruction are obtained by filtering operations, and the sampling grid is uniform. However, all the results herein can be extended to arbitrary Hilbert space settings [12], [13], [14], [15] including finite-dimensional spaces, spaces that are not SI and nonuniform sampling. Our exposition is based on a Hilbert-space interpretation of sampling techniques, and relies on the concepts of bases and projections. This perspective has been motivated in the context of sampling in the excellent review by Unser [11]. Here we consider a similar setting and complement the paper of Unser by surveying further progress made in this area in recent years.

We begin by presenting a broad class of sampling theorems for signals confined to an arbitrary subspace in the presence of non-ideal sampling, and nonlinear distortions. Surprisingly, many types of nonlinearities that are encountered in practice do not pose any technical difficulty and can be completely compensated for. Next, we develop minimax recovery techniques that best approximate an arbitrary smooth input signal. These methods can also be used to reconstruct a signal using a given interpolation kernel that is easy to implement, with only a minor loss in signal quality. To further enhance the quality of the interpolated signal, we discuss fine grid recovery techniques in which the system rate is increased during reconstruction. As we show, the algorithms we develop can all be extended quite naturally to the recovery of random signals. These additional aspects extend the existing sampling framework and incorporate more realistic sampling and interpolation models.

Before proceeding with the detailed development, we note that an additional topic in the context of sampling that has received growing attention recently is that of reconstructing signals that are known to be sparse in some domain. This class of problems underlies the emerging field of compressed sensing [16], [17]. However, this framework has focused primarily on sampling of discrete signals and reconstruction techniques from a finite number of samples, while our interest here is on sampling and reconstructing analog continuous-time signals from uniform samples. Some exceptions are the work in [18], [19], [20], [21] which describe examples of compressed sensing for analog signals, and the work on finite-rate of innovation [22], [23]. In the last section, we very briefly touch on this important area.

## I. Sampling and Reconstruction Setup

The Shannon sampling theorem (also attributed to Nyquist, Whittaker and Kotelnikov) states that a signal $x(t)$ bandlimited to $\pi/T$ can be recovered from its uniform samples at time instants $nT$. Reconstruction is obtained by filtering the samples with a sinc interpolation kernel:

$$x(t) = \frac{1}{T}\sum_{n=-\infty}^{\infty} x(nT)\text{sinc}(t/T-n),$$

TABLE I: Different scenarios treated in this review

| | **Unconstrained Reconstruction** | **Predefined Interpolation Kernel** | **Fine Grid Interpolation** |
|---|---|---|---|
| **Subspace Priors** | Section II-A, II-B | Section II-C | Section II-D |
| **Smoothness Priors** | Section III-A | Section III-B | Section III-C |
| **Stochastic Priors** | Section IV-A | Section IV-B | Section IV-B |

TABLE II: Design objective in each scenario

| | **Unconstrained Reconstruction** | **Predefined Interpolation Kernel** | **Fine Grid Interpolation** |
|---|---|---|---|
| **Subspace Priors** | Perfect reconstruction | Minimum error | Minimum error |
| **Smoothness Priors** | Consistency/minimax | Consistency/minimax | Consistency/minimax |
| **Stochastic Priors** | Mean-squared error (MSE) | MSE | MSE |

where $\text{sinc}(t) = \sin(\pi t)/(\pi t)$. Although widely used, this theorem relies on three fundamental assumptions that are rarely met in practice. First, natural signals are almost never truly bandlimited. Second, the sampling device is usually not ideal, that is, it does not produce the exact signal values at the sampling locations. A common situation is that the ADC integrates the signal, usually over small neighborhoods surrounding the sampling points. Moreover, nonlinear distortions are often introduced during the sampling process. Finally, the use of the sinc kernel for reconstruction is often impractical due to its very slow decay.

To design interpolation methods that are adapted to practical scenarios, there are several issues that need to be properly addressed:

1) The sampling mechanism should be adequately modeled;
2) Relevant prior knowledge about the class of input signals should be taken into account;
3) Limitations should be imposed on the reconstruction algorithm in order to ensure robust and efficient recovery.

In this review we treat each of these three essential components of the sampling scheme. We focus on several models for each of the ingredients, which commonly arise in signal processing, image processing and communication systems. The setups we consider are summarized in Table I, and are detailed in the ensuing subsections. Table II indicates the design objective used in each scenario. As we discuss further below, different priors dictate distinct objectives. For example, when the only information we have about the signal is that it is smooth, then the error cannot be minimized uniformly over all signals, and alternative design strategies are needed.

### *A. Sampling Process*

*1) Linear Distortion:* In the Shannon sampling theorem, $x(t)$ is bandlimited to $\pi/T$ and thus an equivalent strategy is to first filter the signal with a LPF with cut-off $\pi/T$ and then uniformly sample the output. This interpretation is depicted in Fig. 1 with $s(-t) = \text{sinc}(t/T)$ being the impulse response of the LPF. The samples $c[n]$ can be expressed as

$$c[n] = \int_{t=-\infty}^{\infty} x(t)s(t-nT)dt \stackrel{\triangle}{=} \langle x(t), s(t-nT)\rangle, \tag{1}$$

where $\langle y(t), h(t)\rangle$ denotes the $L_2(\mathbb{R})$ inner product between two finite-energy continuous-time real signals. For simplicity, throughout the paper we assume a sampling interval of $T = 1$.

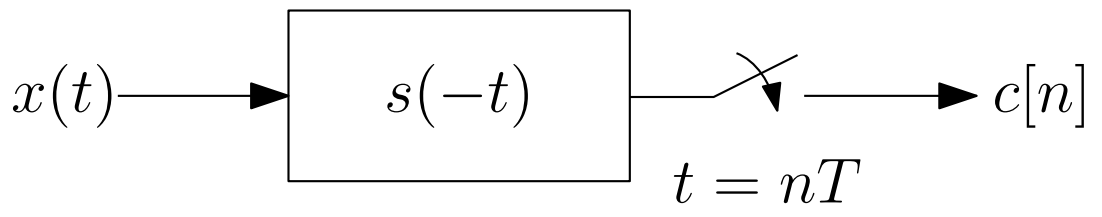

Fig. 1: Shift-invariant sampling. Filtering the signal $x(t)$ prior to taking ideal and uniform samples, can be interpreted as $L_2$ inner-products between $x(t)$ and shifts of $s(t)$. Shannon's framework corresponds to the choice $s(-t) = \text{sinc}(t/T)$.

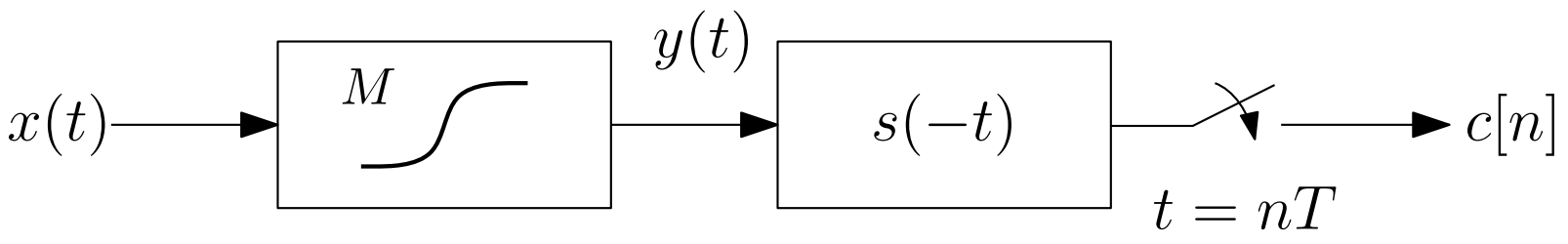

Fig. 2: Nonlinear and shift-invariant sampling. The signal amplitudes $x(t)$ are distorted by the nonlinear mapping $M$ prior to shift-invariant sampling.

In practical applications the sampling is not ideal. Therefore, a more realistic setting is to let $s(t)$ be an arbitrary sampling function. This allows to incorporate imperfections in the ideal sampler into the function $s(t)$ [24], [14], [25], [12]. As an example, typical ADCs average the signal over a small interval rather than outputting pointwise signal values. This distortion can be taken into account by modifying $s(t)$ to include the integration.

*2) Nonlinear Distortion:* A more complicated situation arises when the sampling process includes nonlinear distortions. One simple approach to model nonlinearities is to assume that the signal is distorted by a memoryless, nonlinear, invertible mapping prior to sampling by $s(-t)$, as in Fig. 2. This rather straightforward model is general enough to capture many systems of practical interest. Nonlinearities appear in a variety of setups and applications of digital signal processing including power electronics [26], radiometric photography [27], and CCD image sensors. In some cases, nonlinearity is insinuated deliberately in order to increase the possible dynamic range of the signal while avoiding amplitude clipping, or damage to the ADC [28].

### B. *Signal Priors*

In essence, the Shannon sampling theorem states that if $x(t)$ is known a priori to lie in the space of bandlimited signals, then it can be perfectly recovered from uniformly-spaced ideal samples. Clearly, the question of whether $x(t)$ can be recovered from its samples depends on the prior knowledge we have on the class of input signals. In this review we depart from the traditional bandlimited assumption and discuss signal priors that appear more frequently in signal processing and communication scenarios.

*1) Subspace Priors:* Our first focus is on signal spaces that are SI. A SI subspace $\mathcal{A}$ of $L_2$, is the space of signals that can be expressed as linear combinations of shifts of a generator $a(t)$ [29]:

$$x(t) = \sum_{n=-\infty}^{\infty} b[n]a(t-n), \tag{2}$$

where $b[n]$ is an arbitrary norm-bounded sequence. Note that $b[n]$ does not necessarily correspond to samples of the signal, that is $x(n) \neq b[n]$ in general. Choosing $a(t) = \text{sinc}(t)$ results in the space of $\pi$-bandlimited signals. However, a much broader class of signal spaces can be defined including spline functions [11]. In these cases $a(t)$ can be easier to handle numerically than the sinc function.

A spline $f(t)$ of degree $N$ is a piecewise polynomial with the pieces combined at knots, such that the function is continuously differentiable $N-1$ times. It can be shown that any spline of degree $N$ with knots at the integers can be generated using (2) by a $B$-spline $a(t)$ of degree $N$, which is the function obtained by the $(N+1)$-fold convolution of the unit square

$$b(t) = \begin{cases} 1 & 0 < t < 1; \\ 0 & \text{otherwise.} \end{cases} \tag{3}$$

Signals of the type (2) are also encountered when the analog signal to be sampled originated from a digital source. For example, in communication systems, signals of this form are produced by pulse amplitude modulation. Extensive research in this field has been devoted to design receivers that undo the effect of inter-symbol interference, caused by overlap of the pulses $a(t-n)$. Here we provide a geometric interpretation of this problem, which leads to insight into which classes of signals can be perfectly recovered from their samples. This viewpoint also allows to incorporate various constraints on the reconstruction method.

Although the discussion in this article is limited to SI subspaces, the results we present are valid in more general subspaces as well [12], [13]. In particular, the results can be extended straightforwardly to SI subspaces with multiple generators [30], [31], [21]. In this case, the filters figuring in the sampling and reconstruction are replaced by a bank of filters, and the digital correction is replaced by a multichannel correction system. This allows to treat, for example, signals whose spectrum is contained in several frequency bands.

*2) Smoothness Priors:* Subspace priors are very useful because, as we will see, they often can be utilized to perfectly recover $x(t)$ from its samples. However, in many practical scenarios our knowledge about the signal is much less complete and can only be formulated in very general terms. An assumption prevalent in image and signal processing is that natural signals are smooth in some sense. Here we focus on approaches that quantify the extent of smoothness using the $L_2$ norm $\|Lx(t)\|^2$, where $\|f(t)\|^2 = \langle f(t), f(t)\rangle$, and $L$ is usually chosen as some differential operator. The appeal of these models stems from the fact that they lead to linear recovery procedures. In contrast, smoothness measures such as total variation, result in nonlinear interpolation techniques.

The class of "smooth" signals is much richer than its subspace counterpart. Consequently, it is often impossible to distinguish between one "smooth" signal and another based solely on their samples. In other words, the sampling process inevitably entails information loss. Since perfect recovery cannot be attained in this scenario, we focus on two alternative criteria: consistency (or least-squares) and a worst case (minimax) design.

*3) Stochastic Priors:* The last family we consider in detail is the family of stochastic priors. In this category, the signal $x(t)$ is modeled as a wide-sense stationary (WSS) random process with known power spectral density (PSD), a viewpoint prominent in the field of statistical signal processing. As a design criterion, we focus on minimization of the mean-squared error (MSE) given the signal samples. The theory of sampling random signals has developed in parallel lines to its deterministic counterpart [8]. Interestingly, the stochastic setting leads to reconstruction techniques that are very similar to the methods arising from the smoothness priors. This provides an interesting equivalence between the smoothness operator $L$ and the PSD of $x(t)$ in the random setup. Furthermore, we show that the study of statistical priors also sheds some light on the origin of artifacts, which are commonly encountered

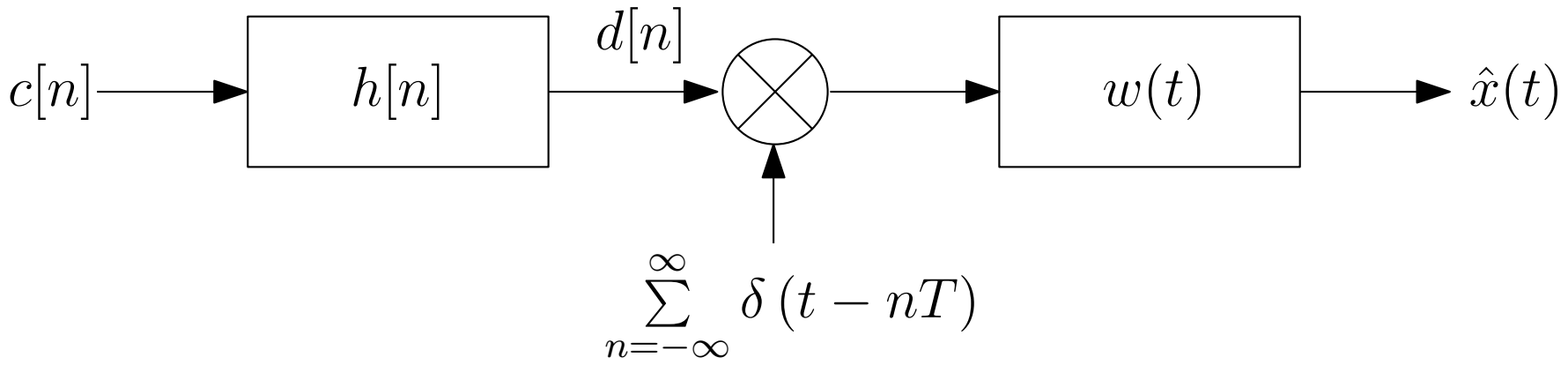

Fig. 3: Reconstruction using a digital compensation filter $h[n]$ and interpolation kernel $w(t)$.

in traditional interpolation methods.

*4) Sparsity Priors:* In the last section we very briefly touch on sparsity priors. This class of functions lead to nonlinear reconstruction algorithms that have a very different structure than the linear interpolation methods in the majority of this paper. Since the treatment of these priors differs substantially from the rest of the review, we only point to several basic recovery techniques and results in this emerging area. A more detailed discussion merits a separate paper.

## C. Reconstruction Methods

For a sampling theorem to be practical, it must take into account constraints that are imposed on the interpolation method. One aspect of the Shannon sampling theorem, which renders it unrealizable, is the use of the sinc interpolation kernel. Due to its slow decay, the evaluation of $x(t)$ at a certain time instant $t_0$, requires using a large number of samples located far away from $t_0$. In many applications, reduction of computational load is achieved by employing much simpler methods, such as linear interpolation. In these cases the sampling scheme should be modified to compensate for the chosen non-ideal kernel.

*1) Unconstrained Reconstruction:* The first setup we consider is unconstrained recovery. Here, we design interpolation methods that are best adapted to the underlying signal prior according to the objectives in Table II, without restricting the reconstruction mechanism. In these scenarios, it is sometimes possible to obtain perfect recovery, as in the Shannon sampling theorem. The unconstrained reconstruction methods under the different scenarios treated in this paper (besides the case in which there are nonlinear distortions) all have a common structure, depicted in Fig. 3. Here $w(t)$ is the impulse response of a continuous-time filter, which serves as the interpolation kernel, while $h[n]$ represents a discrete-time filter used to process the samples prior to reconstruction. Denoting the output of the discrete-time filter by $d[n]$, the input to the filter $w(t)$ is a modulated impulse train $\sum_n d[n]\delta(t-n)$. The filter's output is given by

$$\hat{x}(t) = \sum_{n=-\infty}^{\infty} d[n]w(t-n). \tag{4}$$

Optimal interpolation kernels resulting from such considerations are typically derived in the frequency domain but very often do not admit a closed form in the time domain. This limits the applicability of these recovery techniques to situations in which the kernel needs to be calculated only on a discrete set of points. The discrete Fourier transform (DFT) can be used in such settings to approximate the desired values. Consequently, these methods seem to have been used, for example, in the image processing community only as a means of enlarging an image by an integer factor [32], [33]. More general geometrical transformations, such as rotation, lens distortion correction and scaling by an arbitrary factor, are typically not tackled using these techniques. One way to resolve this problem is to choose the signal prior so as to yield an efficient interpolation algorithm, as done *e.g.*, in [34] in the context of exponential B-splines. Nevertheless, this approach restricts the type of priors that can be handled.

*2) Predefined Kernel:* To overcome the difficulties in implementing the unconstrained solutions, we may resort to a system that uses a predefined interpolation kernel that is easy to implement. In this setup, the only freedom is in the design of the digital correction filter $h[n]$ in Fig. 3, which may be used to compensate for the non-ideal behavior of the pre-specified kernel $w(t)$ [24], [12], [15], [13], [14]. The filter $h[n]$ is selected to optimize a criterion matched to the signal prior.

By restricting the reconstruction to the form (4), we are essentially imposing that the recovered signal $\hat{x}(t)$ lie in the SI space generated by the pre-specified kernel $w(t)$. The class of SI spaces is very general and includes many signal spaces that lead to highly efficient interpolation methods. For example, by appropriate choice of $w(t)$ the family of splines can be described using (4). B-splines have been used for interpolation in the mathematical literature since the pioneering work of Schonberg [35]. In signal processing applications the use of B-splines gained popularity due to the work of Unser et. al. that showed how B-spline interpolation can be implemented efficiently [36], [37]. Interpolation using splines of degree up to $3$ is very common in image processing, due to their ability to efficiently represent smooth signals and the relatively low computational complexity needed for their evaluation at arbitrary locations.

*3) Fine Grid Interpolation:* Constraining the interpolation kernel may lead to severe degradation of the reconstruction. This emphasizes the fundamental tradeoff between performance and implementation considerations. A common way to improve the recovery properties of a reconstruction algorithm is to first upsample the digital signal and then apply some simple interpolation method on the resulting finer grid. This is a widely practiced approach for sampling rate conversion, where usually a rectangular or triangular interpolation filter is used [38].

Under mild conditions on the interpolation kernel, this approach allows to approximate the optimal unconstrained solution arbitrarily well by using a fine enough grid. This, of course, comes at the cost of computational complexity. In practice, it is not the asymptotic behavior that interests us, but rather optimizing the performance for a fixed setup. Thus, given a fixed oversampling factor $K \geq 1$ and an interpolation filter $w(t)$, we would like to design a multirate system that processes the samples $c[n]$ and produces fine-grid expansion coefficients $d[n]$ such that the reconstruction

$$\hat{x}(t) = \sum_{n=-\infty}^{\infty} d[n] w\left(t - \frac{n}{K}\right) \tag{5}$$

well approximates $x(t)$. This setup is depicted in Fig. 4. Besides extending the discussion to general interpolation filters, we also relax the standard assumption that the input signal is bandlimited. Instead, we design a correction system that is best adapted to the prior we have on the input signal.

The interpolation methods corresponding to the different scenarios discussed above are summarized in Table III. The numbers in the table indicate the equation numbers containing the reconstruction formulas. Interestingly, we will see that the solutions share a similar structure. Throughout the article, we emphasize commonalities and equivalence between the different approaches in order to help design the most appropriate filter for a given application. We provide a number of different routes (and formulations) that in many cases lead to the same computational solution, while providing several complementary insights into the problem as well as on the notion of optimality.

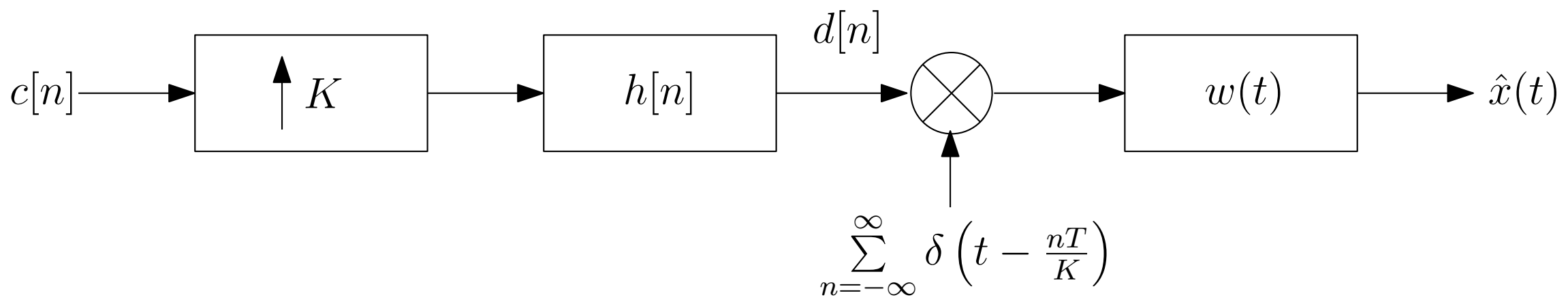

Fig. 4: Fine grid reconstruction using an upsampler followed by a digital compensation filter $h[n]$ and interpolation kernel $w(t)$. The rate of the sequence $d[n]$ is $K$ times larger than that of $c[n]$.

TABLE III: Methods for signal recovery

| | **Unconstrained Reconstruction** | **Predefined Interpolation Kernel** | **Fine Grid Interpolati** |
|---|---|---|---|
| **Subspace Priors** | Linear sampling: (13)<br>Nonlinear distortion: (21), (22), (23) | (24) | (25) |
| **Smoothness Priors** | (28), (29) | Consistent: (33)<br>Minimax: (37) | (39) |
| **Stochastic Priors** | (41), (42) | (45) | (39) |

## II. Subspace Priors

We begin by treating the setting in which the input signal $x(t)$ is known to lie in a given SI subspace. We show that when the reconstruction method is not restricted, these priors allow for perfect recovery of $x(t)$ from its nonideal samples both in the linear setting of Fig. 1 as well as in the presence of nonlinear distortions as in Fig. 2. Specifically, for any sampling function $s(t)$ there are a broad class of subspace priors under which $x(t)$ can be perfectly reconstructed. Conversely, for any given class of functions there are many choices of $s(t)$ that will allow for perfect recovery. These filters only have to satisfy a rather mild requirement. The surprising fact is that these results are valid even when a memoryless, invertible nonlinearity is inserted prior to sampling, as long as the nonlinearity does not vary too fast.

In the second part of this section, we extend the discussion to constrained reconstruction scenarios. In these cases perfect recovery is often impossible, as the restriction narrows down the set of candidate signals which the system can output. However, we will show that it is often possible to produce a reconstruction that minimizes the squared-norm of the error.

Throughout this section $x(t)$ is assumed to lie in a SI subspace $\mathcal{A}$ generated by $a(t)$ (see (2)). In order for $\mathcal{A}$ to be well defined and the corresponding sampling theorems to be stable, the functions $\{a(t-n)\}$ should generate a Riesz basis or a frame [10]. To simplify the exposition we focus throughout on the case in which these functions are linearly independent and therefore form a basis. However, all the results extend easily to the case in which they are linearly dependent. In essence, a Riesz basis is a set of linearly independent vectors that ensures stable expansions, namely a small modification of the expansion coefficients results in a small distortion of the signal (see Box A). In order for $a(t)$ to generate a Riesz basis the continuous-time Fourier transform (CTFT) of $a(t)$ must satisfy

$$\alpha \leq \sum_{k=-\infty}^{\infty} |A(\omega - 2\pi k)|^2 \leq \beta \quad \text{a.e. } \omega, \tag{6}$$

for some constants $\alpha > 0$ and $\beta < \infty$ [39]. The term in the middle of (6) is the discrete-time Fourier transform (DTFT) of the sampled correlation function $r_{aa}[n] = \langle a(t), a(t-n) \rangle$. More details on the CTFT and DTFT are given in Box B. In particular, the functions $\{a(t-n)\}$ form an orthonormal basis if and only if $\alpha = \beta = 1$ in (6).

### A. *Unconstrained Reconstruction with Linear Sampling*

In the setup of Fig. 1 the input signal $x(t)$ is sampled by a set of sampling functions $\{s(t-n)\}$. We denote by $\mathcal{S}$ the space spanned by these sampling functions: any $f(t)$ in $\mathcal{S}$ is of the form

$$f(t) = \sum_{n=-\infty}^{\infty} d[n] s(t-n) \tag{7}$$

for some bounded-norm sequence $d[n]$. We assume throughout that $s(t)$ satisfies the Riesz basis condition (6).

In order to understand what class of signals can be reconstructed from these samples we first observe that knowing the samples $c[n]$ of (1) is equivalent to knowing the orthogonal projection of $x(t)$ onto $\mathcal{S}$, which we denote by $x_{\mathcal{S}}(t) = P_{\mathcal{S}} x(t)$ (see Box C). Indeed,

$$c[n] = \langle x(t), s(t-n) \rangle = \langle x(t), P_{\mathcal{S}} s(t-n) \rangle = \langle P_{\mathcal{S}} x(t), s(t-n) \rangle, \tag{8}$$

where we used the fact that $P_{\mathcal{S}} s(t-n) = s(t-n)$ and $P_{\mathcal{S}}$ is Hermitian. Since the functions $s(t-n)$ span $\mathcal{S}$, and $x_{\mathcal{S}}$ lies in $\mathcal{S}$, it is clear that $x_{\mathcal{S}}$ can be reconstructed from the samples $c[n]$. An immediate consequence is that if $x(t)$ lies in $\mathcal{S}$ so that $x(t) = x_{\mathcal{S}}(t)$, then it can be perfectly recovered.

This geometric interpretation implies that the question of reconstruction from $c[n]$ is equivalent to asking which signals can be recovered from knowledge of their orthogonal projection onto $\mathcal{S}$. At first glance it may seem like only signals in $\mathcal{S}$ may be reconstructed since the projection zeros out any component in $\mathcal{S}^{\perp}$. However, a closer inspection reveals that if we know in advance that $x(t)$ lies in a space $\mathcal{A}$ with suitable properties (which we will define below), then there is a unique vector in $\mathcal{A}$ with the given projection onto $\mathcal{S}$. As depicted in Fig. 5, in this case we can draw a vertical line from the projection until we hit the space $\mathcal{A}$ and in such a way obtain the unique vector in $\mathcal{A}$ that is consistent with the given samples. Evidently, perfect recovery is possible for a broad class of signals beyond those that lie in $\mathcal{S}$.

We next discuss how to recover $x(t)$ explicitly using a discrete-time filter as in Fig. 3. We first note that the orthogonal projection $P_{\mathcal{S}} x(t)$ can be obtained from the samples $c[n]$ by using the scheme in Fig. 3 with $w(t) = s(t)$ and $h[n]$ chosen as the impulse response of the filter with DTFT [40], [11]

$$H(e^{j\omega}) = \frac{1}{\sum_{k=-\infty}^{\infty} |S(\omega - 2\pi k)|^2} = \frac{1}{\phi_{SS}\left(e^{j\omega}\right)}, \tag{9}$$

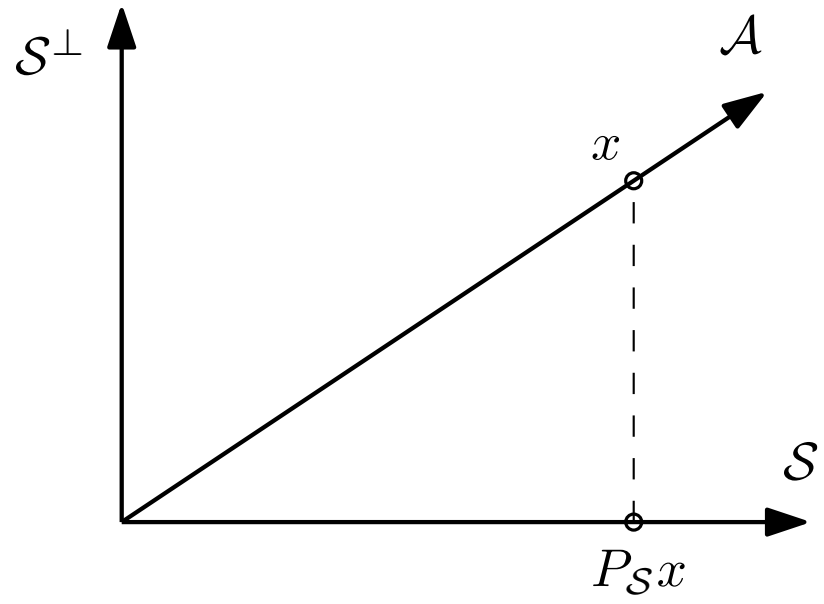

Fig. 5: A unique vector in $\mathcal{A}$ which is consistent with the samples in $\mathcal{S}$ can be recovered from the known samples.

where $S(\omega)$ is the CTFT of $s(t)$,

$$\phi_{SA}\left(e^{j\omega}\right) \triangleq \sum_{k=-\infty}^{\infty} S^*(\omega - 2\pi k)A(\omega - 2\pi k), \tag{10}$$

and $(\cdot)^*$ denotes the complex conjugate. Here $A(\omega)$ is the CTFT of an arbitrary function $a(t)$. The function $\phi_{SA}(e^{j\omega})$ is the DTFT of the sampled cross-correlation sequence $r_{sa} = \langle s(t), a(t-n)\rangle$ (See Box B). Note that the Riesz basis condition (6) guarantees that (9) is well defined. Efficient implementation of (9), and the filters we introduce in the sequel, is possible in spline spaces, based on the results of [11], [36], [37].

To show that the output of the resulting system is $P_{\mathcal{S}}x(t)$ note that if $x(t)$ is in $\mathcal{S}^\perp$, then the output will be zero since in this case $c[n]$ is the zero sequence. This follows from the fact the inner product of $x(t)$ with any signal in $\mathcal{S}$ is zero. On the other hand, if $x(t) \in \mathcal{S}$, then from (7) we can write $x(t) = \sum_n b[n]s(t-n)$ for some sequence $b[n]$. Using the Fourier relations given in Box B it follows that

$$C(e^{j\omega}) = B(e^{j\omega}) \sum_{k=-\infty}^{\infty} |S(\omega - 2\pi k)|^2 = B(e^{j\omega})\phi_{SS}\left(e^{j\omega}\right). \tag{11}$$

Therefore, $d[n] = b[n]$ and $\hat{x}(t) = x(t)$. Consequently, if $x(t)$ lies in $\mathcal{S}$ to begin with, then this scheme will ensure perfect reconstruction. If in addition $s(t)$ satisfies the partition of unity property, that is $\sum_n s(t-n) = 1$ for all $t$, then it can be shown that by selecting the sampling period $T$ sufficiently small, any input signal that is norm bounded can be approximated as close as desired by this approach [11].

The denominator in (9) is the DTFT of the sampled correlation function $r_{ss}[n] = \langle s(t), s(t-n)\rangle$. Therefore, if the functions $\{s(t-n)\}$ form an orthonormal basis, then $r_{ss}[n] = \delta[n]$ and $H(e^{j\omega}) = 1$. In this case no pre-processing of the samples is necessary prior to reconstruction. This is precisely the setting of the Shannon sampling theorem: it is easy to verify that the functions $s(t-n) = \text{sinc}(t-n)$ form an orthonormal basis [41], [11].

To extend recovery beyond the space $\mathcal{S}$, suppose that $x(t)$ lies in a known subspace $\mathcal{A}$. Clearly in order to be able to reconstruct $x(t)$ from the given samples we need that $\mathcal{A}$ and $\mathcal{S}^\perp$ intersect only at zero, since any non-zero signal $y(t)$ in the intersection of $\mathcal{A}$ and $\mathcal{S}^\perp$ will yield zero samples and therefore cannot be recovered. Throughout, we say that two spaces are disjoint if they intersect only at zero. Intuitively, we also need $\mathcal{A}$ and $\mathcal{S}$ to have the same number of degrees of freedom. These requirements can be made precise by assuming a direct sum condition $L_2 = \mathcal{A} \oplus \mathcal{S}^\perp$ where $\oplus$ denotes a sum of two subspaces that intersect only at the zero vector. This implies that $\mathcal{A}$ and $\mathcal{S}^\perp$ are disjoint, and together span the space of $L_2$ signals. In the SI setting this condition translates into a simple requirement on the CTFT of the generators $a(t), s(t)$ of $\mathcal{A}, \mathcal{S}$ [42]:

$$\left|\phi_{SA}\left(e^{j\omega}\right)\right| > \alpha, \tag{12}$$

for some constant $\alpha > 0$, where $\phi_{SA}\left(e^{j\omega}\right)$ is defined by (10). Under this condition, reconstruction can be obtained by choosing $w(t) = a(t)$ and [24], [15], [13], [14], [12]

$$H(e^{j\omega}) = \frac{1}{\phi_{SA}\left(e^{j\omega}\right)}. \tag{13}$$

When $\mathcal{A} = \mathcal{S}$, the filter (13) coincides with (9).

To see that (13) ensures perfect recovery for signals in $\mathcal{A}$, note that any $x(t) \in \mathcal{A}$ can be written as $x(t) = \sum_n b[n]a(t-n)$. Using the relations in Box B it can be shown that the sequence of samples will have a DTFT given by

$$C(e^{j\omega}) = B(e^{j\omega})\phi_{SA}(e^{j\omega}), \tag{14}$$

from which the result follows. In addition, for any $x(t) \in \mathcal{S}^\perp$ we have immediately that $\hat{x}(t) = 0$ since $c[n]$ will be the zero sequence. Consequently, the overall system implements an oblique projection $E_{\mathcal{A}\mathcal{S}^\perp}$ with range space $\mathcal{A}$ and null space $\mathcal{S}^\perp$ [43], [39] (see Box C). Indeed, this is the unique operator satisfying $E_{\mathcal{A}\mathcal{S}^\perp}x(t) = x(t)$ for any $x(t)$ in $\mathcal{A}$, and $E_{\mathcal{A}\mathcal{S}^\perp}x(t) = 0$ for all $x(t)$ in $\mathcal{S}^\perp$.

It is also interesting to interpret the proposed sampling scheme as a basis expansion. Since any signal in $\mathcal{A}$ can be recovered from the corrected samples $d[n] = c[n] * h[n]$ via $x(t) = \sum_n d[n]a(t-n)$, we may view this sequence as the coefficients in a basis expansion. To obtain the corresponding basis we note that by combining the effects of the sampler $s(t)$ and the correction filter $h[n]$ of (13), the sequence of samples can be equivalently expressed as $d[n] = \langle x(t), v(t-n)\rangle$ where $v(t) = \sum_n h[n]s(t-n)$. In the Fourier domain,

$$V(\omega) = H(e^{j\omega})S(\omega). \tag{15}$$

Therefore, we conclude that any $x(t) \in \mathcal{A}$ can be written as

$$x(t) = \sum_{n=-\infty}^{\infty} \langle x(t), v(t-n)\rangle a(t-n). \tag{16}$$

It can be easily verified that the functions $\{v(t-n)\}$ form a Riesz basis for $\mathcal{S}$, and $\langle v(t-n), a(t-m)\rangle = \delta_{mn}$ where $\delta_{mn} = 1$ if $m = n$ and $0$ otherwise. Therefore, these functions are the oblique dual basis of $\{a(t-n)\}$ in $\mathcal{S}$ [14], [15], [13], [42], [44], [31] (see Box A). When $\mathcal{A} = \mathcal{S}$, we recover the conventional dual basis functions. In this case $\{v(t-n)\}$ forms a basis for $\mathcal{S}$ that is dual to the original basis $\{s(t-n)\}$: $\langle v(t-n), s(t-m)\rangle = \delta_{nm}$. This provides a concrete method for constructing a dual of a given basis $\{a(t-n)\}$ in any subspace $\mathcal{S}$ satisfying the direct sum condition $L_2 = \mathcal{A} \oplus \mathcal{S}^\perp$.

To conclude our discussion so far, we have seen that a signal $x(t)$ in a SI subspace $\mathcal{A}$ generated by $a(t)$, can be reconstructed from its generalized samples in Fig. 1 using any choice of $s(t)$ for which (12) is satisfied. Thus for a given SI space, there is a broad variety of sampling filters we can select from. By choosing the functions appropriately, a variety of interesting sampling theorems can be formulated, such as pointwise sampling of non-bandlimited signals, bandlimited sampling of nonbandlimited functions, and many more. An example is given below.

Despite the fact that any sampling function $s(t)$ satisfying (12) can be used to sample $x(t)$ in the space $\mathcal{A}$ generated by $a(t)$, in the presence of noise out of $\mathcal{A}$, the choice of sampling kernel will effect the reconstructed signal. More specifically, we have seen that the output of Fig. 3 with $w(t) = a(t)$ and $h[n]$ given by (13) is equal to the oblique projection $x_E(t) = E_{\mathcal{A}\mathcal{S}^\perp}x(t)$. When $x(t) \in \mathcal{A}$, we have $x_E(t) = x(t)$ for any choice of $\mathcal{S}^\perp$, or equivalently any sampling function $s(t)$ in Fig. 3 satisfying (12). However, if $x(t)$ does not lie entirely in $\mathcal{A}$, for example due to noise, then different functions $s(t)$ will result in different approximations $x_E(t) \in \mathcal{A}$. A natural question is: Given an interpolation kernel $a(t)$, which choice of sampling function $s(t)$ will lead to a reconstruction $\hat{x}(t)$ that is closest to $x(t)$? If we measure the error using the squared-norm $\|\hat{x}(t) - x(t)\|^2$, then the choice $s(t) = a(t)$ minimizes the error. This follows from the projection theorem which states that the orthogonal projection onto $\mathcal{A}$ is the closest vector in $\mathcal{A}$ to an arbitrary input $x(t)$ [45]:

$$\arg \min_{v(t)\in\mathcal{A}} \|x(t) - v(t)\|^2 = P_{\mathcal{A}}x(t). \tag{17}$$

Therefore, since using a kernel $a(t)$ will lead to an interpolation $\hat{x}(t)$ in $\mathcal{A}$ irrespective of $h[n]$, the smallest error will result when $\hat{x}(t) = P_{\mathcal{A}}x(t)$. The orthogonal projection can be achieved only if the sampling function $s(t)$ generates $\mathcal{A}$. In addition, in contrast with the orthogonal projection, an oblique projection can increase the norm of the noise at the input (see Box C). In practice, however, we may prefer other choices that are easier to implement at the expense of a slight increase in error [24], [12].

We conclude this subsection with a non-intuitive example in which a signal that is not bandlimited is filtered with a LPF prior to sampling, and still can be perfectly reconstructed from the resulting samples.

Consider a signal $x(t)$ formed by exciting an RC circuit with a modulated impulse train $\sum_n d[n]\delta(t-n)$, as shown in Fig. 6(a). The impulse response of the RC circuit is known to be $a(t) = \tau^{-1}\exp\{-t/\tau\}u(t)$, where $u(t)$ is the unit step function and $\tau = RC$ is the time constant. Therefore

$$x(t) = \frac{1}{\tau}\sum_{n=-\infty}^{\infty} d[n]\exp\{-(t-n)/\tau\}u(t-n). \tag{18}$$

Clearly, $x(t)$ is not bandlimited. Now, suppose that $x(t)$ is filtered by an ideal LPF $s(t) = \mathrm{sinc}(t)$ and then sampled at times $t = n$ to obtain the sequence $c[n]$. The signal $x(t)$ and its samples are depicted in Fig. 6(b). Intuitively, there seems to be information loss in the sampling process since the entire frequency content of $x(t)$ outside $[-\pi, \pi]$ is zeroed out, as shown in Fig. 6(c). However, it is easily verified that condition (12) is satisfied in this setup and therefore perfect recovery is possible. The digital correction filter (13) in this case can be shown to be

$$h[n] = \begin{cases} 1 & n = 0; \\ \frac{\tau}{n}(-1)^n & n \neq 0. \end{cases} \tag{19}$$

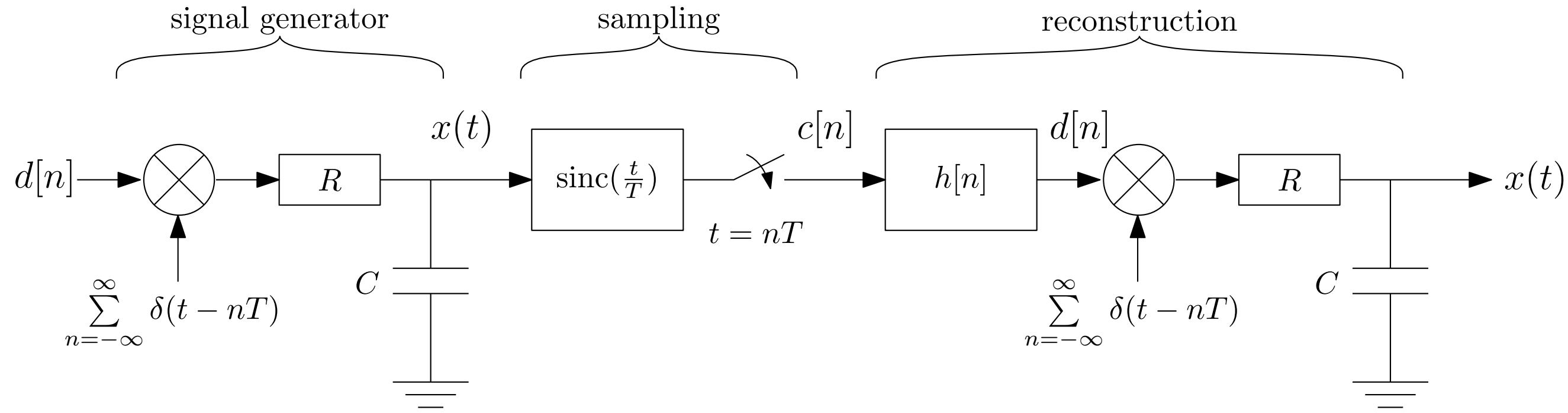

(a) Sampling and reconstruction scheme.

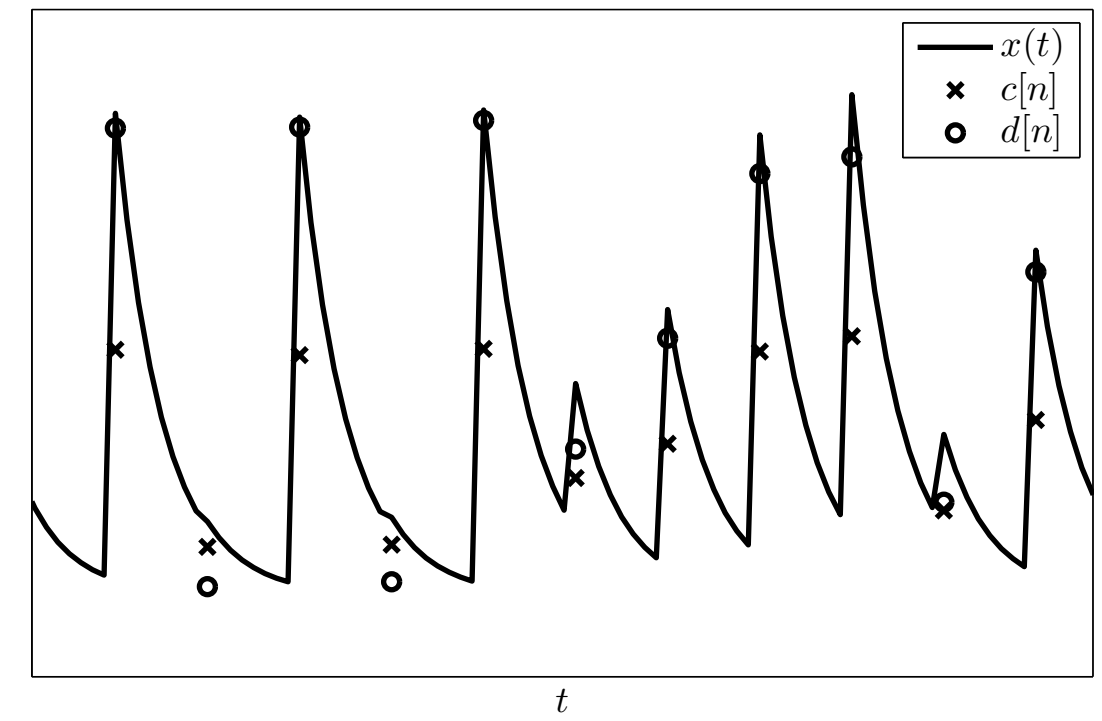

(b) The signal, its samples and expansion coefficients.

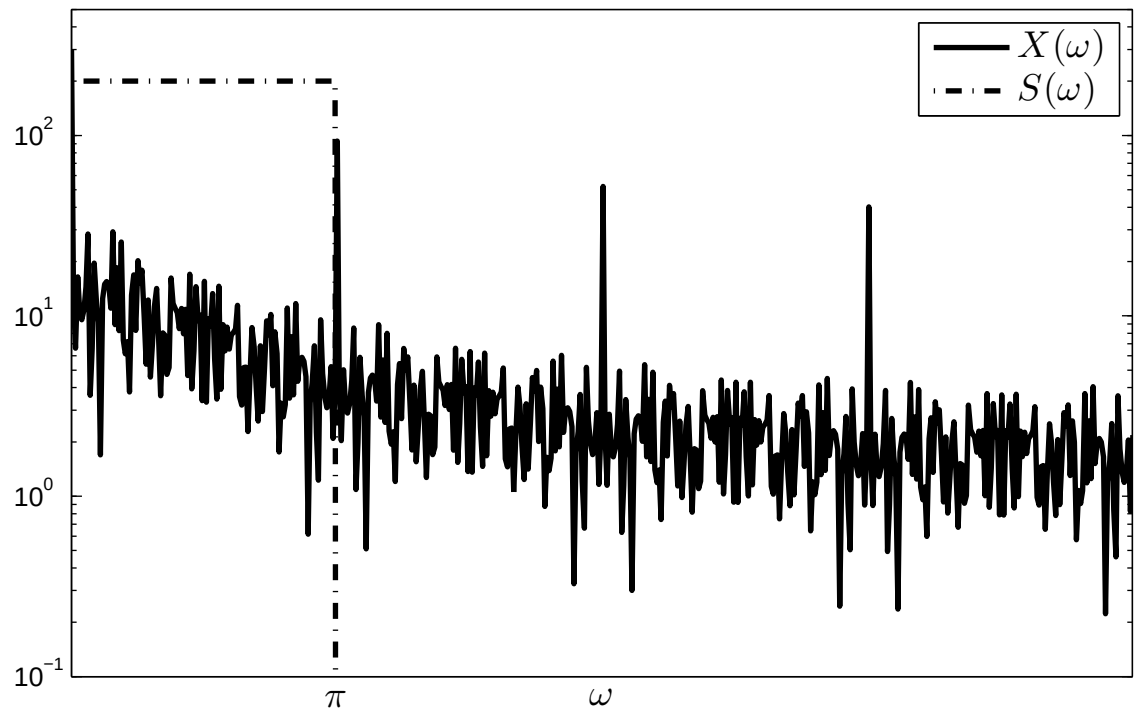

(c) Frequency content of the signal and sampling filter.

Fig. 6: A non-bandlimited signal $x(t)$, formed by exciting an RC-circuit with a modulated impulse train, is sampled after passing through an ideal LPF and then perfectly reconstructed by re-exciting an identical RC-circuit with an impulse train modulated by a digitally filtered version of the samples. (a) Sampling and reconstruction setup. (b) The signal $x(t)$ and its samples $c[n]$. (c) The signal $X(\omega)$ and the sampling filter $S(\omega)$. Perfect recovery is possible despite the fact that a large portion of the frequency content is lost due to the filtering operation.

Thus, to reconstruct $x(t)$ we need to excite an identical RC circuit with an impulse train modulated by the sequence $d[n] = h[n] * c[n]$. The entire sampling-reconstruction setup is depicted in Fig. 6(a).

### *B. Unconstrained Reconstruction with Nonlinear Distortion*

Suppose now that as in the previous section $x(t)$ lies in a subspace $\mathcal{A}$ and (12) is satisfied. However, prior to sampling by $s(-t)$ the signal is distorted by a memoryless, nonlinear and invertible mapping $M(x)$ as in Fig. 2. A naive approach to recover the signal $x(t)$ from its samples is to first apply $M^{-1}$ to the sample sequence $c[n]$, leading to a sequence $d[n]$, and then reconstruct $x(t)$ from the samples $d[n]$ using standard reconstruction techniques [46]. However, if the samples $c[n]$ are not ideal, namely are not pointwise evaluations of $x(t)$, then this approach is suboptimal in general.

A surprising result developed in [47] is that if the nonlinearity is invertible and does not change too fast, then it does not introduce theoretical difficulties. More specifically, under the same direct sum condition (12) we had in the linear sampling case, and assuming that the derivative of the nonlinearity is appropriately bounded, there is a unique signal $x(t)$ with the given samples $c[n]$. Therefore, it is enough to seek a recovery $\hat{x}(t)$ that is consistent in the sense that it yields the samples $c[n]$ after it is reinjected into the system: $\int_{-\infty}^{\infty} s(t-n)M\left(\hat{x}(t)\right)dt = c[n]$.

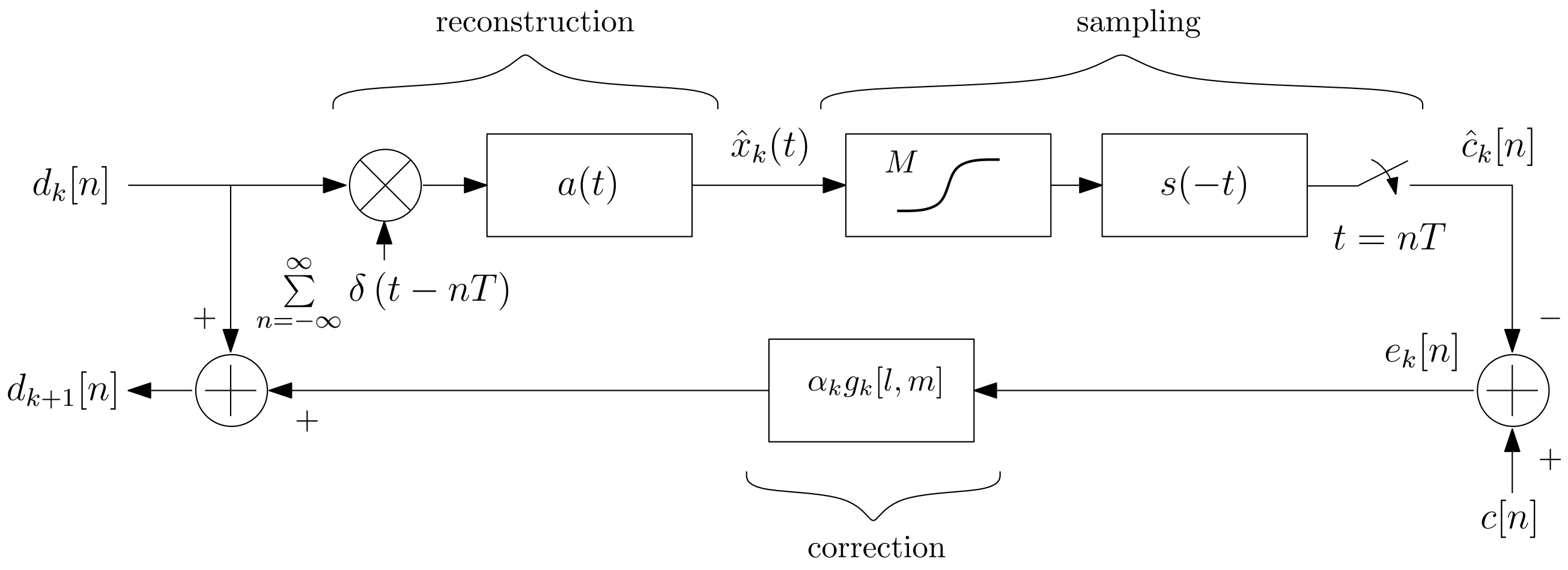

Fig. 7: One iteration of the nonlinear recovery algorithm. The expansion coefficients at the $k$th iteration, $d_k[n]$, are used to synthesize an estimate of the signal, $\hat{x}_k(t)$. This estimate goes through the sampling process to produce the corresponding samples $\hat{c}_k[n]$. The error with respect to the measured samples, $e_k[n] = c[k] - \hat{c}_k[n]$, is then used to update the estimate.

Any such signal must be equal to $x(t)$ due to the uniqueness property. This result is important as it allows to reformulate the recovery problem in terms of minimizing the error in the samples.

Since $x(t) \in \mathcal{A}$, we can write $\hat{x}(t) = \sum_n d[n]a(t-n)$ for some sequence $d[n]$. Thus, our problem reduces to finding a sequence $d[n]$ which minimizes the consistency cost function

$$f(d) = \|c[n] - \hat{c}[n]\|_{\ell_2}. \tag{20}$$

Here $\|b[n]\|_{\ell_2}$ is the $\ell_2$-norm of the sequence $b[n]$, and

$$\hat{c}[n] = \int_{-\infty}^{\infty} s(t-n) M\left(\sum_{m=-\infty}^{\infty} d[m]a(t-m)\right) dt, \tag{21}$$

are the estimated samples based on our current guess of $x(t)$. Clearly the minimal value of $f(d)$ is 0. Since $M$ is nonlinear, the cost function (20) is in general non-convex. Therefore optimization algorithms for minimizing (20) might trap a stationary point, and not the global minimum which we seek. Surprisingly, it can be shown [47] that under the direct sum condition and appropriate bounds on the derivative of $M$, (20) has a unique stationary point which is equal to the global minimum. Therefore, any algorithm designed to trap a stationary point automatically leads to perfect recovery. This is despite the fact that the objective (20) is not convex. In Fig. 7 we show a block diagram of an iterative approach which is derived by applying a Newton method on (20). This same algorithm can also be obtained from an approximate projection onto convex sets strategy, and a linearization approach; see [47] for more details.

At each iteration, the algorithm of Fig. 7 works as follows. Denote by $d_k[n]$ the expansion coefficients at the $k$th iteration so that $\hat{x}_k(t) = \sum_n d_k[n]a(t-n)$. Then $d_{k+1}[n]$ is calculated as

$$d_{k+1}[n] = d_k[n] + \alpha_k \sum_{m=-\infty}^{\infty} g_k[n, m] e_k[m], \tag{22}$$

where $\alpha_k$ is the step size, $e_k[m] = c[m] - \hat{c}_k[m]$ is the error-in-samples sequence with $\hat{c}_k[n]$ denoting the approximate samples at stage $k$ obtained via (21) with $d[n] = d_k[n]$, and $g_k[l,m]$ is a linear system which is the inverse of

$$h_k[l,m] = \int_{-\infty}^{\infty} s(t-l)M'\left(\sum_{n=-\infty}^{\infty} d_k[n]a(t-n)\right) a(t-m)dt. \tag{23}$$

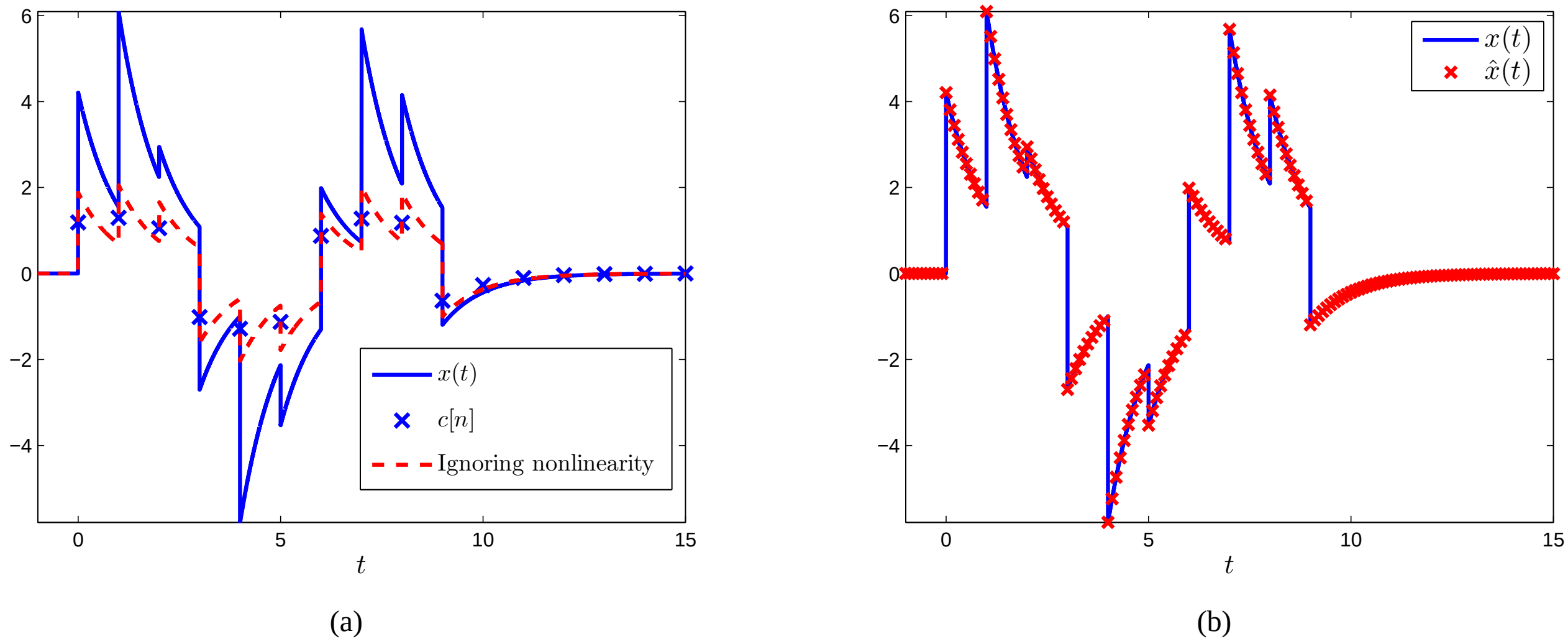

Fig. 8: A signal $x(t)$ lying in a shift-invariant space was linearly sampled after passing through a memoryless nonlinear system. (a) Ignoring the nonlinear distortion and filtering the samples $c[n]$ with the filter $H(e^{j\omega})$ of (13), leads to poor reconstruction. (b) The algorithm presented here leads to perfect recovery of $x(t)$.

Here $M'$ denotes the derivative of $M$. Note that $h_k[l,m]$ is not SI in general and therefore it cannot be inverted in the frequency domain to obtain $g_k[l,m]$. In practice, though, one usually analyzes a finite set of samples $c[n], 0 \leq n \leq N-1$. Assuming that $c[n] = 0$ outside this range, the matrix $\{g_k[l,m]\}$ for $0 \leq l, m \leq N-1$ can be obtained by inverting the corresponding matrix $\{h_k[l,m]\}$.

We now demonstrate the effectiveness of the algorithm in a scenario similar to that of Fig. 6. Specifically, suppose that, as in Fig. 6, $x(t)$ is known to be of the form (18), and we are given the samples $c[n] = \int_{n-1}^{n} \arctan\left(x(t)\right) dt$. This corresponds to using a nonlinear mapping $M(x) = \arctan(x)$ and a sampling filter with impulse response equal to a rectangular window of length 1. The signal and its samples are depicted in Fig. 8(a). Evidently, the samples $c[n]$ constitute a rather poor representation of the signal. Consequently, if one ignores the nonlinearity and uses the techniques developed in the previous section, that is filtering with $H(e^{j\omega})$ of (13), then the reconstruction error is large (dotted line). In Fig. 8(b) we show the result of applying the algorithm presented here, which leads to perfect reconstruction of $x(t)$ from the nonideal samples $c[n]$.

### *C. Constrained Reconstruction*

Up until now we specified the sampling process but did not restrict the reconstruction or interpolation kernel $w(t)$ in Fig. 3. In many applications this kernel is fixed in advance due to implementation issues. For example, in image processing applications kernels with small supports are often used. These include nearest neighbor, bilinear, bicubic, Lanczos and splines. The interpolation kernel $w(t)$ can also represent the pixel shape of an image display.

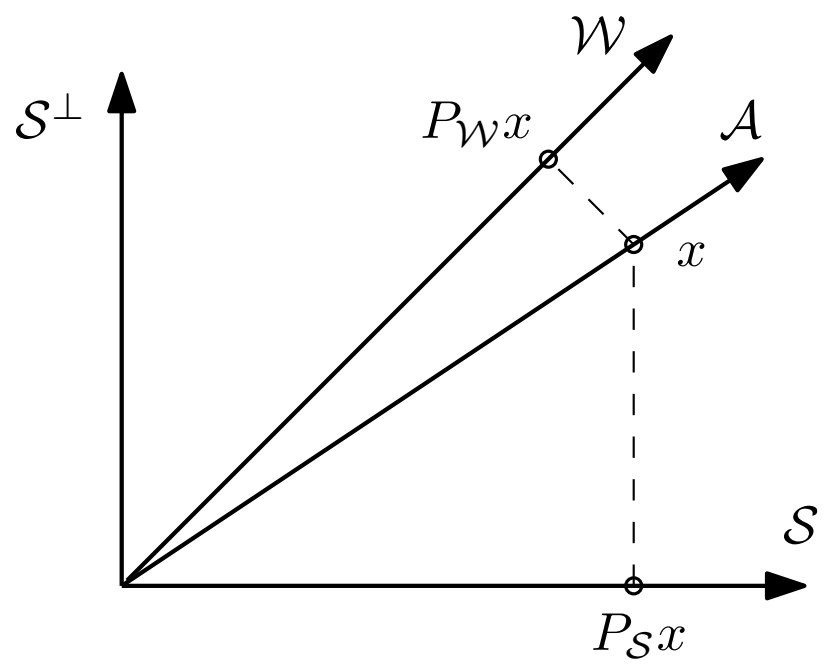

Fig. 9: The signal $x(t) \in \mathcal{A}$ can be recovered from the samples $c[n]$, allowing to compute its orthogonal projection onto $\mathcal{W}$.

In order to obtain stable reconstruction, we concentrate in the sequel on cases in which $w(t)$ satisfies the Riesz basis condition (6). In particular, it can be easily shown that $B$-splines all satisfy this requirement.

Given a sampling function $s(-t)$ and a fixed interpolation kernel $w(t)$ an important question is how to design the digital filter $h[n]$ in Fig. 3 so that the output $\hat{x}(t)$ is a good approximation of the input signal $x(t)$ in some sense. Clearly, $\hat{x}(t)$ will always lie in the space $\mathcal{W}$, spanned by the generator $w(t)$. This is because for every choice of the sequence $d[n]$, $\hat{x}(t)$ has the form $\hat{x}(t) = \sum_n d[n]w(t-n)$. Therefore, if $x(t)$ does not lie in $\mathcal{W}$ to begin with, then $\hat{x}(t)$ cannot be equal $x(t)$. Since $\hat{x}(t)$ is constrained to lie in $\mathcal{W}$, it follows from the projection theorem (17) that the minimal error approximation to $x(t)$ is obtained when $\hat{x}(t) = P_{\mathcal{W}}x(t)$. The question is whether this solution can be generated from the samples $c[n]$. In general, the answer is negative without sufficient prior knowledge on the signal [12]. However, when $x(t)$ lies in a subspace satisfying (12), $P_{\mathcal{W}}x(t)$ can be obtained by filtering the sample sequence with

$$H\left(e^{j\omega}\right) = \frac{\phi_{WA}\left(e^{j\omega}\right)}{\phi_{SA}\left(e^{j\omega}\right)\phi_{WW}\left(e^{j\omega}\right)}, \tag{24}$$

where $\phi_{WA}\left(e^{j\omega}\right)$, $\phi_{SA}\left(e^{j\omega}\right)$ and $\phi_{WW}\left(e^{j\omega}\right)$ are as in (10) with the corresponding substitution of the filters $W(\omega)$, $A(\omega)$ and $S(\omega)$. In this case, the output of the system of Fig. 3 is given by $P_{\mathcal{W}}E_{\mathcal{A}\mathcal{S}^\perp}x(t)$ [12]. Consequently, if $x(t) \in \mathcal{A}$, then $E_{\mathcal{A}\mathcal{S}^\perp}x(t) = x(t)$ and the minimal squared-error approximation $P_{\mathcal{W}}x(t)$ is achieved.

To understand this result geometrically, note that we have already seen in the previous subsection that under the direct sum condition, any vector $x(t) \in \mathcal{A}$ can be recovered from the samples $c[n]$. This is illustrated in Fig. 5. Here, however, we are constrained to obtain a solution in $\mathcal{W}$. But, since we can determine $x(t)$, we can also compute $P_{\mathcal{W}}x(t)$, which is the minimal squared-error approximation in $\mathcal{W}$. This is shown in Fig. 9.

### *D. Dense Grid Recovery*

The situation in which $x(t)$ can be completely determined from its samples but cannot be reproduced by the system is somewhat frustrating. Moreover, the error caused by restricting the recovered signal to lie in $\mathcal{W}$ may be very large if $\mathcal{W}$ is substantially different from $\mathcal{A}$. One way to bridge the gap between the unconstrained and constrained recovery techniques is to increase the interpolation rate, namely produce a reconstruction of the form $\hat{x}(t) = \sum_{=-\infty}^{\infty} d[n]w(t-n/K)$, for some integer $K > 1$, as depicted in Fig. 4. This strategy is legitimate as we are still using a predefined interpolation kernel $w(t)$, which may be easy to implement. Thus, we effectively introduce a tradeoff between complexity and performance.

The motivation for this approach can be understood from a geometric viewpoint. As we increase the interpolation rate $K$, the reconstruction space $\mathcal{W}$ spanned by the functions $\{w(t-n/K)\}$ becomes "larger" and consequently "closer" to $\mathcal{A}$. In some cases, there exists a factor $K$ for which $\mathcal{W}$ contains $\mathcal{A}$, thus recovering the possibility of perfect reconstruction.

In order for the reconstruction to be stable, we focus on the case in which the functions $\{w(t-n/K)\}$ form a Riesz basis. This requirement is satisfied if and only if there exists constants $0<\alpha\leq\beta<\infty$ such that $\alpha\leq\sum_{l=-\infty}^{\infty}|W(\omega-2\pi lK)|^2\leq\beta$ is satisfied almost everywhere.

Similarly to the setting in which $K=1$, it can be shown that when $x(t)$ is in $\mathcal{A}$, the minimal squared error solution $\hat{x}(t)=P_{\mathcal{W}}x(t)$ can be attained with the system depicted in Fig. 4. The frequency response of the correction filter $h[n]$, which operates on the up-sampled data, is given by

$$H(e^{j\omega})=\sum_{m=0}^{K-1}\frac{\phi_{W_{\mathrm{s}}A_{\mathrm{s}}}\left(e^{j(\omega+2\pi m/K)}\right)}{\phi_{SA}\left(e^{jK\omega}\right)\phi_{W_{\mathrm{s}}W_{\mathrm{s}}}\left(e^{j(\omega+2\pi m/K)}\right)},\tag{25}$$

where $W_{\mathrm{s}}(\omega)=W(K\omega)$, $A_{\mathrm{s}}(\omega)=A(K\omega)$, and $\phi_{SA}(e^{j\omega})$ is defined in (10). The dense grid recovery scheme of Fig. 4 can also be implemented using polyphase filters, which in some cases may lead to a simpler implementation [48].

## III. Smoothness Priors

Up until now we considered the setting in which the input signal $x(t)$ is constrained to a subspace. We now treat a more general and less restrictive formulation of the sampling problem in which our prior knowledge on the signal is that it is smooth in some sense. Here we model the extent of smoothness of $x(t)$ as the $L_2$ signal-norm at the output of a continuous-time filter $L(\omega)$ with $x(t)$ as its input. In practice, $L(\omega)$ is often chosen to be a first or second order derivative in order to constrain the solution to be smooth and nonoscillating, *i.e.*, $L(\omega)=a_0+a_1j\omega+a_2(j\omega)^2+\cdots$ for some constants $a_n$. Another common choice is the filter $L(\omega)=(a_0^2+\omega^2)^\gamma$ with some parameter $\gamma$. We denote the output norm as $\|Lx(t)\|$. For simplicity, we assume that $L(\omega)>\alpha>0$ almost everywhere for some $\alpha$, although the results extend to the non-invertible case as well.

Unlike subspace priors, a one-to-one correspondence between smooth signals and their sampled version does not exist since smoothness is a far less restrictive constraint than confining the signal to a subspace. Perfect recovery, or even error-norm minimization, is therefore impossible. Indeed, it can be shown that there is no single choice of $\hat{x}(t)$ that minimizes $\|\hat{x}(t)-x(t)\|^2$ over all smooth signals $x(t)$, even when $\hat{x}(t)$ is constrained to lie in a subspace $\mathcal{W}$. This is because the sample sequence $c[n]$ is no longer sufficient to determine the orthogonal projection $P_{\mathcal{W}}\hat{x}(t)$ [12]. Therefore, below, we focus on alternative approaches for designing the reconstruction system.

### A. Unconstrained Reconstruction

To approximate $x(t)$ from its samples, based solely on the knowledge that it is smooth, we consider two design techniques. The first consists of finding the smoothest signal which gives rise to the measured samples $c[n]$ [24].

The second is a minimax strategy in which the system is designed to yield the best approximation for the worst-case signal among smooth inputs that are consistent with the samples [12].

*1) Smoothest Approximation:* In this approach we require that the reconstructed signal $\hat{x}(t)$ is smooth and consistent with the samples. The consistency requirement means that $\hat{x}(t)$ should yield the same samples $c[n]$ when reinjected into the system:

$$\langle \hat{x}(t), s(t-n) \rangle = c[n] = \langle x(t), s(t-n) \rangle \quad \text{for all } n. \tag{26}$$

The simplest strategy to produce a consistent smooth reconstruction is to minimize the smoothness $\|Lx(t)\|$ subject to the consistency requirement:

$$\hat{x}(t) = \arg\min_{x(t)} \|Lx(t)\| \text{ subject to } S\{x(t)\} = \boldsymbol{c}. \tag{27}$$

The notation $S\{x(t)\}$ denotes the sequence of samples $\langle s(t-n), x(t) \rangle$ and $\boldsymbol{c}$ stands for the sequence $\{c[n]\}$. It can be shown that the solution to (27) has the form of Fig. 3 where now the reconstruction kernel is

$$\tilde{W}(\omega) = \frac{S(\omega)}{|L(\omega)|^2}, \tag{28}$$

and

$$H(e^{j\omega}) = \frac{1}{\phi_{S\tilde{W}}(e^{j\omega})}. \tag{29}$$

In the previous section, we have seen that the filter (29) corresponds to the choice leading to perfect reconstruction for signals $x(t) \in \tilde{\mathcal{W}}$ (see (12)). Thus, this approach can be viewed as first determining the optimal space given by (28), and then finding the unique signal in $\tilde{\mathcal{W}}$ that is consistent with the given samples.

As a special case, we may choose to produce the minimal norm consistent reconstruction $\hat{x}(t)$ by letting $L$ be the identity operator $I$. This leads to $\tilde{w}(t) = s(t)$ and $H(e^{j\omega})$ is then given by (9). Consequently, $\hat{x}(t)$ is the orthogonal projection onto the sampling space, $\hat{x}(t) = P_{\mathcal{S}}x(t)$. This can also be seen by noting that any reconstruction $\hat{x}(t)$ which yields the samples $c[n]$ has the form $\hat{x}(t) = P_{\mathcal{S}}x(t) + v(t)$ where $v(t)$ is an arbitrary vector in $\mathcal{S}^{\perp}$. The minimal norm approximation corresponds to the choice $v(t) = 0$.

*2) Minimax Recovery:* The reconstruction error $\|\hat{x}(t) - x(t)\|^2$ of any recovery method depends on the unknown original signal $x(t)$. This renders comparison between interpolation methods complicated. Indeed, one algorithm may be better than another for certain input signals and worse for others. The next approach we discuss is based on optimizing the squared-error performance for the worst input signal.

The prior information we have can be used to construct a set $\mathcal{V}$ of all possible input signals:

$$\mathcal{V} = \{x(t) : S\{x(t)\} = \boldsymbol{c}, \|Lx(t)\| \leq U\}, \tag{30}$$

were $U > 0$ is some finite constant. The set consists of signals that are consistent with the samples and are relatively smooth (with respect to the weighted norm $\|Lx(t)\|$). We now seek the reconstruction that minimizes

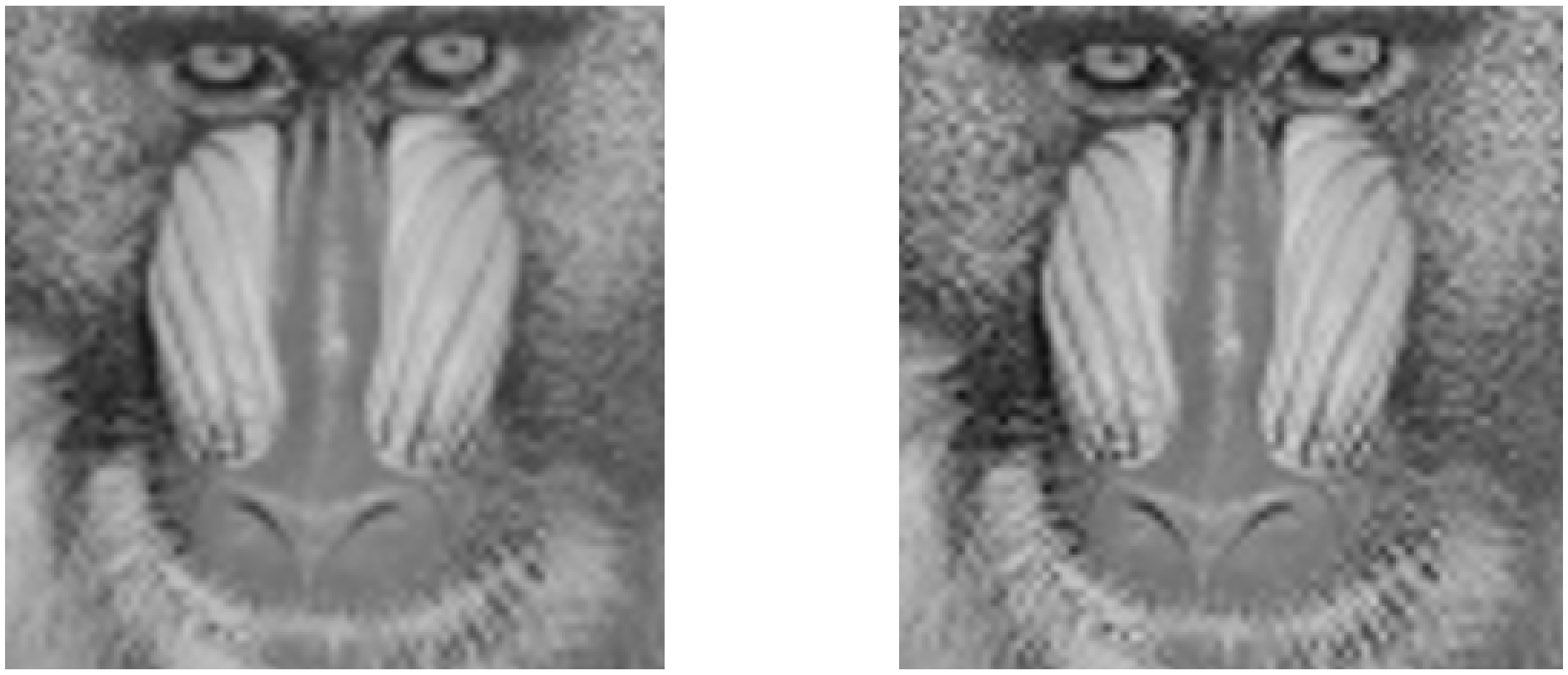

(a) Bicubic (b) Minimax

Fig. 10: Mandrill image rescaling: down-sampling by a factor of $3$ using a rectangular sampling filter followed by upsampling back to the original dimensions using two interpolation methods. (a) The bicubic interpolation kernel leads to a blurry reconstruction with PSNR of $24.18$dB. (b) The minimax method leads to a sharper reconstruction with PSNR of $24.39$dB.

the worst-case error over $\mathcal{V}$:

$$\min_{\hat{x}(t)} \max_{x(t)\in\mathcal{V}} \|\hat{x}(t) - x(t)\|^2. \tag{31}$$

It can be shown that the optimal solution does not depend on the constant $U$. Furthermore, the minimax solution interestingly coincides with the smoothest approximation method, that is, the optimal interpolation kernel and digital compensation filter are given by (28) and (29) respectively.

Although the two approaches we discussed are equivalent in the unrestricted setting, the minimax strategy allows more flexibility in incorporating constraints on the reconstruction, as we show in the next subsection. Furthermore, it tends to outperform the consistency approach when further restrictions are imposed as we will demonstrate via several examples.

Figure 10 compares the minimax approach with bicubic interpolation in the context of image enlargement. The regularization operator was taken to be $L(\boldsymbol{\omega}) = \left((0.1\pi)^2 + \|\boldsymbol{\omega}\|^2\right)^{1.3}$, where $\boldsymbol{\omega}$ denotes the 2D frequency vector. In this example minimax recovery is superior to the commonly used bicubic method in terms of peak signal to noise ratio (PSNR), defined as $\mathrm{PSNR} = 10\log_{10}(255^2/\mathrm{MSE})$ with MSE denoting the empirical squared-error average over all pixel values. In terms of visual quality, the minimax reconstruction is sharper and contains enhanced textures.

### *B. Constrained Reconstruction*

We next treat the problem of approximating $x(t)$ from its samples $c[n]$ using a pre-specified interpolation kernel $w(t)$. Similar to the unrestricted scenario, the two main approaches in this setup are consistent reconstruction [24], [11], [14], [15], [13] and minimax recovery [12], [49]. However, here the solutions no longer coincide. These methods can both be interpreted in terms of projections onto the spaces $\mathcal{W}$ and $\mathcal{S}$ that figure in the problem setting.

Whereas the first approach leads to an oblique projection, the second strategy involves orthogonal projections, rendering this solution more robust to noise [50], [51].

*1) Consistent Reconstruction:* In order to incorporate the constraint on the interpolation kernel, we extend (27) to include the restriction $x(t) \in \mathcal{W}$:

$$\hat{x}(t) = \arg\min_{x(t)} \|Lx(t)\| \text{ subject to } S\{x(t)\} = \boldsymbol{c},\ x(t) \in \mathcal{W}. \tag{32}$$

Recall that under the direct sum condition (12) with $\mathcal{W}$ playing the role of $\mathcal{A}$, there is a unique signal in $\mathcal{W}$ satisfying $S\{x(t)\} = \boldsymbol{c}$, which is equal to the oblique projection $E_{\mathcal{W}\mathcal{S}^\perp}x(t)$. Since there is only one signal in the constraint set of problem (32), the smoothness measure in the objective does not play a role. The oblique projection can be obtained by processing the samples $c[n]$ using the filter

$$H(e^{j\omega}) = \frac{1}{\phi_{SW}\left(e^{j\omega}\right)}. \tag{33}$$

Comparing with (13), we see that this is precisely the filter that yields perfect recovery when we know that $x(t) \in \mathcal{W}$. When the direct sum condition is not satisfied, there can be several consistent solutions so that the objective in (32) is needed in order to select one output among all possibilities [52], [53].

*2) Minimax Recovery:* A drawback of the consistency approach is that the fact that $x(t)$ and $\hat{x}(t)$ yield the same samples does not necessarily imply that $\hat{x}(t)$ is close to $x(t)$. Indeed, for an input $x(t)$ not in $\mathcal{W}$, the norm of the resulting reconstruction error $\hat{x}(t) - x(t)$ can be made arbitrarily large, if $\mathcal{S}$ is close to $\mathcal{W}^\perp$. Furthermore, as we have seen, the consistency method essentially ignores the smoothness prior.

In order to directly control the reconstruction error $\|\hat{x}(t) - x(t)\|^2$, we may modify the minimax strategy of the previous subsection to include the restriction $x(t) \in \mathcal{W}$. Therefore, our minimax design criterion is now:

$$\min_{\hat{x}(t)\in\mathcal{W}} \max_{x(t)\in\mathcal{V}} \|\hat{x}(t) - x(t)\|^2, \tag{34}$$

where $\mathcal{V}$ is the set of smooth consistent signals given by (30).

It turns out that the criterion (34) is too conservative and, for example, in the case in which $L$ is the identity operator $L = I$ it results in the trivial solution $\hat{x}(t) = 0$ [12]. To counterbalance the conservative behavior of the minimax approach, instead of minimizing the worst-case squared-norm error, we consider minimizing the worst-case *regret* [54]. The regret is defined as the difference between the squared-norm error and the smallest possible error that can be achieved with a reconstruction in $\mathcal{W}$, namely $\|P_{\mathcal{W}^\perp}x(t)\|^2$. This error is attained when $\hat{x}(t) = P_{\mathcal{W}}x(t)$, which in general cannot be computed from the sequence of samples $c[n]$. Since the regret depends in general on $x(t)$, it cannot be minimized for all $x(t)$. Instead we consider minimizing the worst-case regret over all possible signals $x(t)$ that are consistent with the given samples, which results in the problem

$$\min_{\hat{x}(t)\in\mathcal{W}} \max_{x(t)\in\mathcal{V}} \left\{ \|\hat{x}(t) - x(t)\|^2 - \|P_{\mathcal{W}^\perp}x(t)\|^2 \right\}, \tag{35}$$

with $\mathcal{V}$ given by (30). Expressing $x(t)$ as $x(t) = P_{\mathcal{W}}x(t) + P_{\mathcal{W}^\perp}x(t)$ and recalling that $\hat{x}(t) \in \mathcal{W}$ it is easy to see

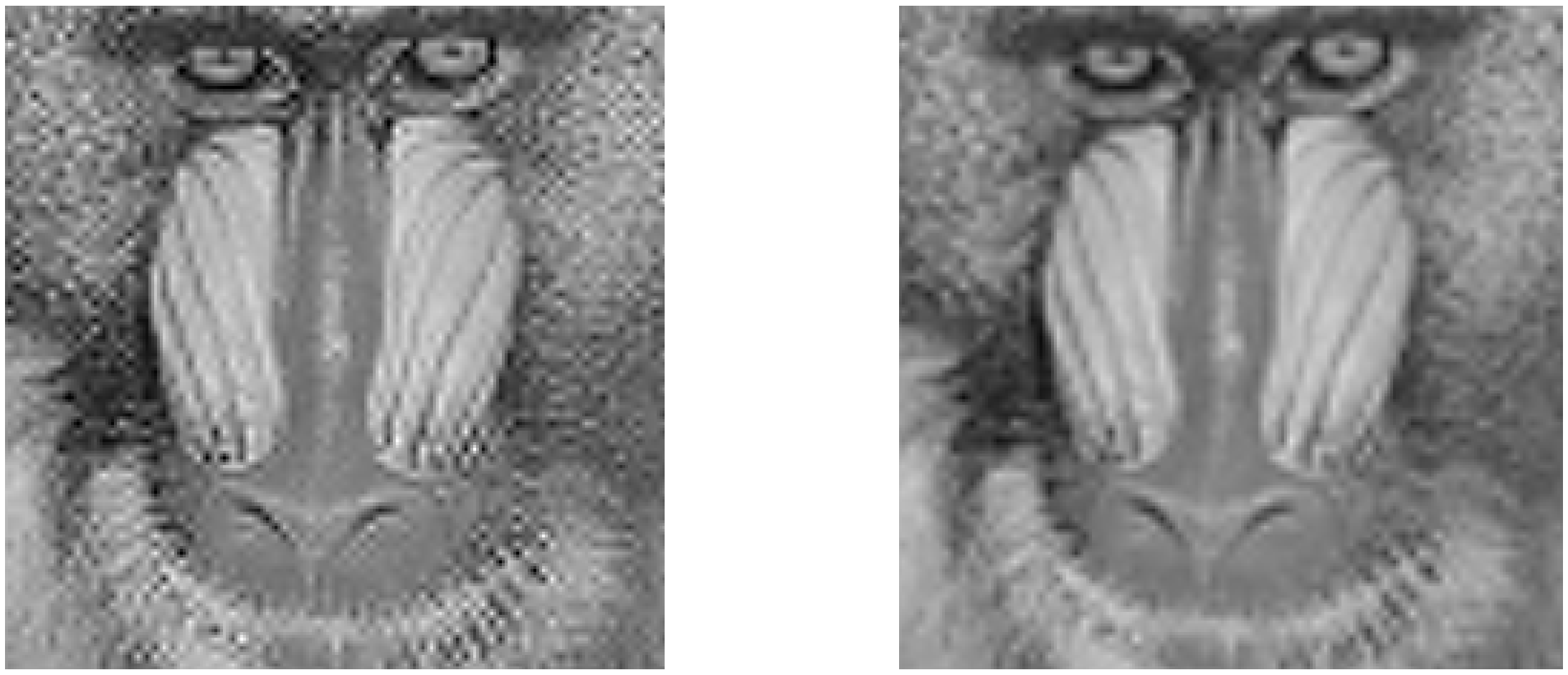

(a) Consistent. (b) Minimax regret.

Fig. 11: Mandrill image rescaling: down-sampling by a factor of $3$ using a rectangular sampling filter followed by upsampling back to the original dimensions using a triangular interpolation kernel via the consistent and minimax regret methods. (a) The consistent approach over-enhances the high frequencies and results in a PSNR of $22.51$dB. (b) The minimax regret method leads to a smoother reconstruction with PSNR of $23.69$dB.

that (35) is equivalent to

$$\min_{\hat{x}(t)\in\mathcal{W}} \max_{x(t)\in\mathcal{V}} \|\hat{x}(t) - P_{\mathcal{W}} x(t)\|^2 . \tag{36}$$

The solution to (36) can be shown to be the projection onto $\mathcal{W}$ of the unconstrained minimax recovery given by (28) and (29). The reconstructed signal $\hat{x}(t)$ is obtained by digitally filtering the samples $c[n]$ with the filter

$$H(e^{j\omega}) = \frac{\phi_{W\tilde{W}}\left(e^{j\omega}\right)}{\phi_{S\tilde{W}}\left(e^{j\omega}\right)\phi_{WW}\left(e^{j\omega}\right)}, \tag{37}$$

where $\tilde{W}(\omega)$ is given by (28).

In Fig. 11 we demonstrate the difference between the consistent and minimax-regret methods in an image-enlargement task. The setup is the same as that of Fig. 10 only now the reconstruction filter is constrained to be a triangular kernel corresponding to linear interpolation. It can be seen that the error of the minimax regret recovery is only $0.7$dB less than the unconstrained minimax shown in Fig. 10. The consistent approach, on the other hand, is much worse both in terms of PSNR and in terms of visual quality. Its tendency to over-enhance high frequencies stems from the fact that it ignores the smoothness prior.

Many of the interesting properties of the minimax-regret recovery (37) can be best understood by examining the case where our only prior on the signal is that it is norm-bounded, that is, when $L$ is the identity operator $L = I$. This choice of $L$ was thoroughly investigated in [12]. Setting $L(\omega) = 1$ in (37), the correction filter becomes

$$H(e^{j\omega}) = \frac{\phi_{WS}\left(e^{j\omega}\right)}{\phi_{SS}\left(e^{j\omega}\right)\phi_{WW}\left(e^{j\omega}\right)}, \tag{38}$$

since from (28), $\tilde{w}(t) = s(t)$. Applying the Cauchy-Schwartz inequality to the numerator of (38) and to the denominator of (33), it is easy to see that the magnitude of the minimax regret filter (38) is smaller than that of the consistent filter (33) at all frequencies. This property renders the minimax regret approach more resistant to noise in the samples $c[n]$, since perturbations in $\hat{x}(t)$ caused by errors in $c[n]$ are always smaller in the minimax regret method than in the consistent approach.

Apart from robustness to digital noise, which takes place after the signal was sampled, the minimax regret method is also more resistant to perturbations in the continuous-time signal $x(t)$. To see this note that the minimax regret reconstruction is given by $\hat{x}(t) = P_{\mathcal{W}} P_{\mathcal{S}} x(t)$. Thus, the norm of $\hat{x}(t)$ is necessarily bounded by that of $x(t)$. Furthermore, it is easy to show that the resulting reconstruction error is always bounded by twice the norm of $x(t)$: $\|\hat{x}(t) - x(t)\|^2 \leq 2\|x(t)\|^2$. In contrast, the consistent recovery is given by the oblique projection $\hat{x}(t) = E_{\mathcal{W}\mathcal{S}^\perp} x(t)$, which may increase the norm of $x(t)$. Consequently, the error of the consistent reconstruction can, in some cases, grow without bound.

In Fig. 12 we illustrate the minimax regret reconstruction geometrically for the case $L = I$. We have seen already that knowing the samples $c[n]$ is equivalent to knowing $x_{\mathcal{S}}(t) = P_{\mathcal{S}} x(t)$. In addition, our recovery is constrained to lie in the space $\mathcal{W}$. As illustrated in the figure, the minimax regret solution is a robust recovery scheme in which the signal is first orthogonally projected onto the sampling space, and then onto the reconstruction space.

When $x(t)$ is known to lie in $\mathcal{S}$, it follows from the previous section that the minimal error can be obtained by using (24) with $A(\omega) = S(\omega)$. The resulting filter coincides with the minimax regret filter of (38). Consequently, the regret approach minimizes the squared-error over all $x(t) \in \mathcal{S}$.

An interesting feature of the minimax regret solution is that it does not depend on the norm bound $U$. Therefore, $\hat{x}(t) = P_{\mathcal{W}} P_{\mathcal{S}} x(t)$ minimizes the worst-case regret error over all bounded inputs $x(t)$, regardless of the norm of $x(t)$. Furthermore, the regret recovery method does not require the direct-sum condition $L_2 = \mathcal{W} \oplus \mathcal{S}^\perp$, which is necessary in the development of the unique consistent approach.

In [12] tight bounds on the error resulting from each of the methods are developed and compared. We omit the technical details here and only summarize the main conclusions. We first recall that if we know a priori that $x(t)$ lies in a subspace $\mathcal{A}$ such that $L_2 = \mathcal{A} \oplus \mathcal{S}^\perp$, then the filter (24) will yield the minimal error approximation of $x(t)$ and therefore is optimal in the squared-norm sense. When $\mathcal{A} = \mathcal{S}$ this strategy reduces to the minimax regret method, while if $\mathcal{A} = \mathcal{W}$, then we obtain the consistent reconstruction. When no prior subspace knowledge is given, the regret approach is preferable if the spaces $\mathcal{S}$ and $\mathcal{W}$ are sufficiently far apart, or if $x(t)$ has enough energy in $\mathcal{S}$. These results are intuitive as illustrated geometrically in Fig. 12. In Fig. 12(a) we depict the consistent and regret reconstruction when $\mathcal{W}$ is far from $\mathcal{S}$. As can be seen in the figure, in this case the error resulting from the consistent solution is large with respect to the regret approximation error. In Fig. 12(b), $\mathcal{W}$ and $\mathcal{S}$ are close, and the errors have roughly the same magnitude.

### *C. Minimax Dense Grid Reconstruction*

We now extend the minimax regret approach to the dense-grid recovery setup of Fig. 4, in which the interpolation is performed using a predefined kernel $w(t)$ on a grid with $1/K$ spacings. To treat this scenario within the minimax-regret framework, we need to solve (34) with the appropriate reconstruction space, namely $\mathcal{W} = \text{span}\{w(t - n/K)\}$.

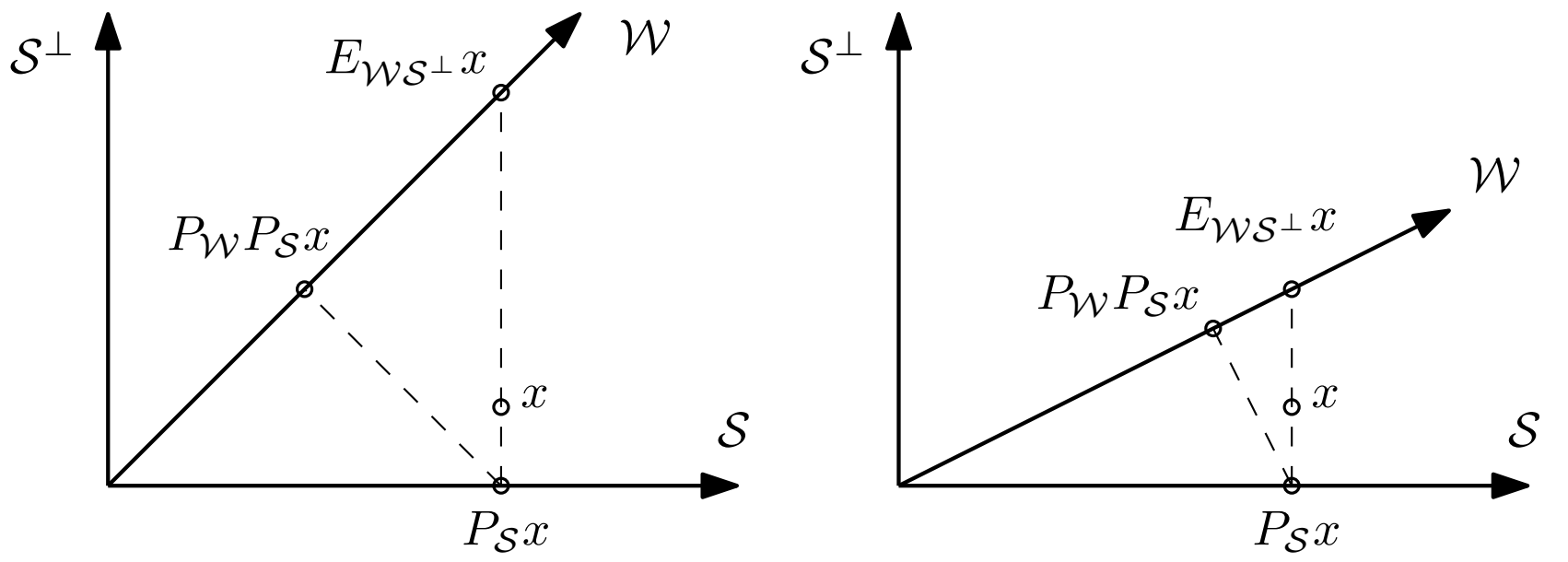

Fig. 12: Comparison of minimax regret reconstruction ($P_\mathcal{W}P_\mathcal{S}x(t)$) and consistent reconstruction ($E_{\mathcal{W}\mathcal{S}^\perp}x(t)$) for two different choices of $\mathcal{W}$. (a) The minimax strategy is preferable when $\mathcal{W}$ is 'far' from $\mathcal{S}$. (b) Both methods lead to errors on the same order of magnitude when $\mathcal{W}$ is 'close' to $\mathcal{S}$.

The corresponding correction filter can be shown to be

$$H(e^{j\omega}) = \sum_{m=0}^{K-1} \frac{\phi_{W_\text{s}\tilde{W}_\text{s}}\left(e^{j(\omega+2\pi m/K)}\right)}{\phi_{S\tilde{W}}\left(e^{jK\omega}\right)\phi_{W_\text{s}W_\text{s}}\left(e^{j(\omega+2\pi m/K)}\right)}, \tag{39}$$

where $W_\text{s}(\omega) = W(K\omega)$ and $\tilde{W}_\text{s}(\omega) = \tilde{W}(K\omega)$ with $\tilde{W}(\omega)$ of (28).

To understand the necessity of fine grid interpolation, note that there is no analytic expression for the optimal unconstrained kernel (28) in the time domain. In rate conversion situations, where the output rate is an integer multiple of the input rate, the kernel $w(t)$ needs to be calculated only on a discrete set of points. This is because $\hat{x}(n/K) = \sum_m d[m]w(n/K - m)$, where $K$ is the oversampling factor. To approximate the sequence $\{w(n/K)\}$ on a finite set of indices, one can sample the expression $\sum_l W(\omega - 2\pi lK)$ on a finite set of frequencies and apply the inverse DFT. However, if $\hat{x}(t)$ must be evaluated at arbitrary locations, then this method cannot be used.

In the previous subsection we have seen that this problem can be tackled by using a predefined interpolation kernel for which a formula exists. An alternative approach is to first evaluate the optimal kernel (28) on a dense grid, and then use nearest neighbor or linear interpolation to obtain the values at the desired locations. This is referred to as first and second order approximation [38]. It is easy to show that these approaches correspond to using the high-rate scheme of Fig. 4 with the digital correction filter

$$h_{ap}[n] = g(n/K), \tag{40}$$

and with a rectangular or triangular interpolation kernel $w(t)$. Here $G(\omega) = H(e^{j\omega})\tilde{W}(\omega)$ with $H(e^{j\omega})$ of (29) and $\tilde{W}(\omega)$ given by (28). Note, therefore, that this method does not take into account the non-optimal interpolation to be performed in the second stage. This is in contrast with the dense grid approach presented here, where the correction filter (39) explicitly depends on $w(t)$. Filter (39) shapes the spectrum of the up-sampled sequence in a way that partially compensates for the non-optimal kernel to follow.

In Fig. 13 we compare the minimax-regret dense-grid reconstruction approach and first-order approximation to the unconstrained filter. To emphasize the differences, we used both methods to enlarge an image by an irrational factor $\pi/e$. It is clearly seen that the first-order approximation approach produces blurry reconstruction whereas

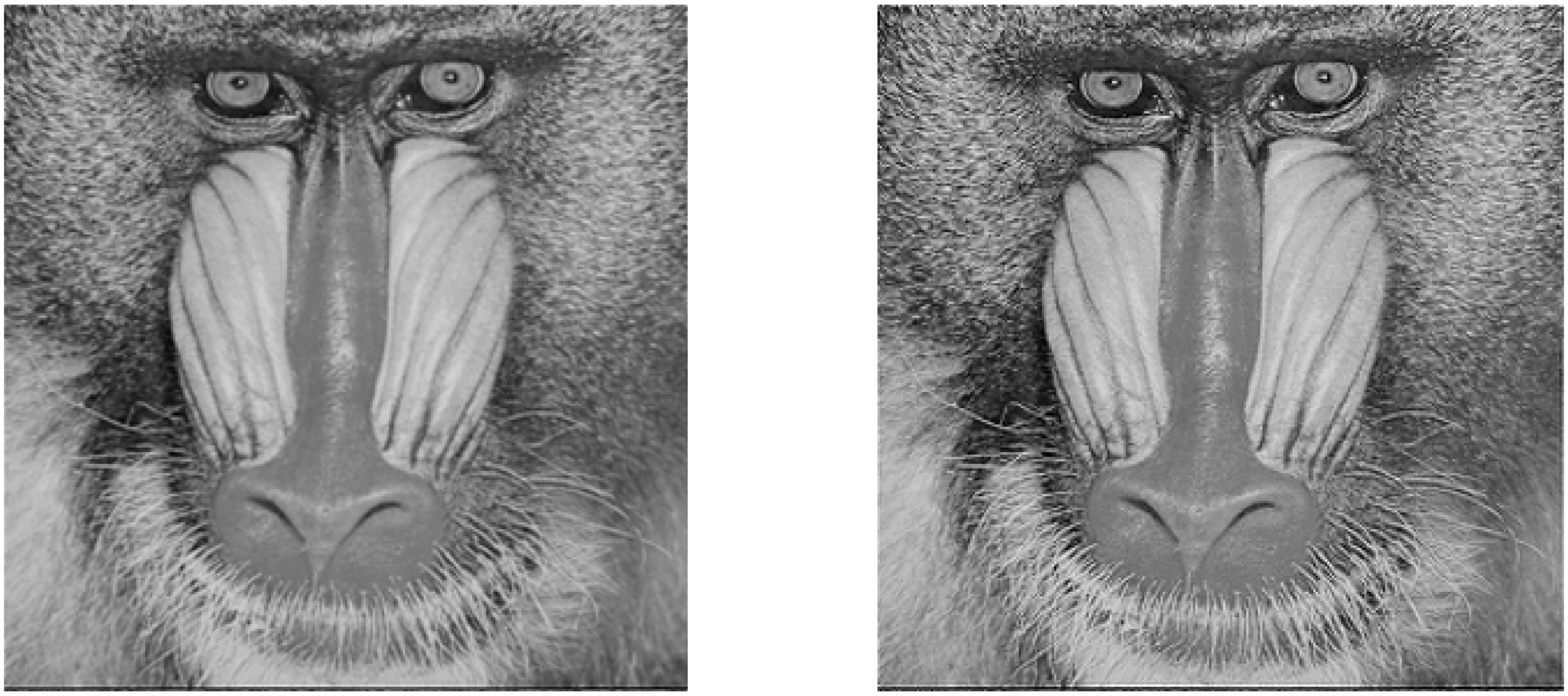

(a) First order approximation of the minimax approach. (b) Dense grid minimax regret.

Fig. 13: Comparison of first order approximation to the minimax method and dense grid minimax regret. In both methods the image is up-sampled by a factor of $K = 2$ and digitally filtered. Then, linear interpolation (triangular kernel) is used to calculate the gray-level values at the desired locations. (a) First order approximation to minimax regret (40). (b) Dense grid minimax regret (39).

in the minimax method the edges are sharp and the textures are better preserved.

## IV. Stochastic Priors

In this section we treat signal priors of stochastic nature. Specifically, the input $x(t)$ is modeled as a WSS random process having PSD $\Lambda_{xx}(\omega)$. Our goal is to linearly estimate $x(t)$ given the samples $c[n]$. As we will see, the schemes resulting from these considerations have strong connections to the minimax methods of the previous section. In addition, this viewpoint also offers a nice explanation to reconstruction artifacts, frequently encountered in applications.

### *A. Unconstrained Reconstruction*

We first examine constrained-free reconstruction. In the deterministic setting with smoothness prior we could not minimize the squared-error $\|\hat{x}(t) - x(t)\|^2$ for all smooth $x(t)$, and therefore discussed a minimax method instead. In contrast, in the stochastic setting we can use the PSD $\Lambda_{xx}(\omega)$ of $x(t)$ in order to minimize the MSE $E[|x(t) - \hat{x}(t)|^2]$ for every $t$, which depends only on the statistics of $x(t)$ and not on the signal itself.

Our approach is to minimize the MSE by linear processing of the samples $c[n]$. As opposed to the common Wiener filtering problem, where both the input and output are either continuous- or discrete-time signals, here we are interested in estimating the continuous-time signal $x(t)$ based on equidistant samples of $y(t) = x(t) * s(-t)$. Consequently, we refer to this as the hybrid Wiener filtering problem.

The reconstruction $\hat{x}(t)$ minimizing the MSE can be implemented by the block diagram in Fig. 3 with the interpolation kernel [55], [56], [33], [57]

$$W(\omega) = S(\omega)\Lambda_{xx}(\omega), \tag{41}$$

and digital correction filter

$$H(e^{j\omega}) = \frac{1}{\sum_{k=-\infty}^{\infty} |S(\omega - 2\pi k)|^2 \Lambda_{xx}(\omega - 2\pi k)}. \tag{42}$$

It is interesting to observe that (41) and (42) are identical to (28) and (29) with $\Lambda_{xx}(\omega) = |L(\omega)|^{-2}$. Therefore, the smoothness operator in the deterministic case corresponds to the whitening filter of the input $x(t)$ in the stochastic setting.

### B. Constrained Reconstruction

We now treat a more practical constrained setting, in which the interpolation filter is fixed in advance. Unfortunately, in this case, for a general given interpolation kernel, there is no digital correction filter that can minimize the MSE for every $t$ [50]. In fact, the filter minimizing the MSE at a certain time instant $t_0$ also minimizes the MSE at times $\{t_0 + n\}$ for all integer $n$, but not over the whole continuum. Therefore, error measures other than pointwise MSE must be considered. Before treating the problem of choosing an appropriate criterion, we first discuss how this time dependence phenomenon is related to artifacts commonly encountered in certain interpolation methods.

The signal $x(t)$ in our setup is assumed to be WSS and, consequently, the sequence of samples $c[n]$ is a discrete WSS random process, as is the output $d[n]$ of the digital correction filter in Fig. 4. The reconstruction $\hat{x}(t)$ is formed by modulating the shifts of the kernel $w(t)$ by the WSS discrete-time process $d[n]$. Assuming that the PSD of $d[n]$ is positive everywhere, signals of this type are not stationary unless $w(t)$ is $\pi$–bandlimited [48]. Generally, $\hat{x}(t)$ will be a cyclostationary process. In practice, the interpolation kernels in use have a finite (and usually small) support, and are therefore not bandlimited. In these cases, the periodic correlation in $\hat{x}(t)$ often degrades the quality of the reconstruction, as subjectively perceived by the visual or auditory system.

Note that although natural signals are rarely stationary to begin with, it is still relevant to study how an interpolation algorithm reacts to stationary signals. In fact, if an interpolation scheme outputs a cyclostationary signal when fed with a stationary input, then it will commonly produce reconstructions with degraded subjective quality also when applied to real world signals, as demonstrated in Fig. 14. However, periodic structure in a recovered signal is not necessarily related to MSE. For example, the optimal unrestricted kernel (41) is usually not bandlimited and thus leads to periodic structure in $\hat{x}(t)$.

The nonstationary behavior of $\hat{x}(t)$ is the reason why the pointwise MSE can not be minimized for every $t$ in general. Two alternative error measures that have been proposed are the sampling-period-average-MSE and the projected MSE.

The sampling-period-average-MSE utilizes the periodicity of the MSE, and integrates it over one period [58],

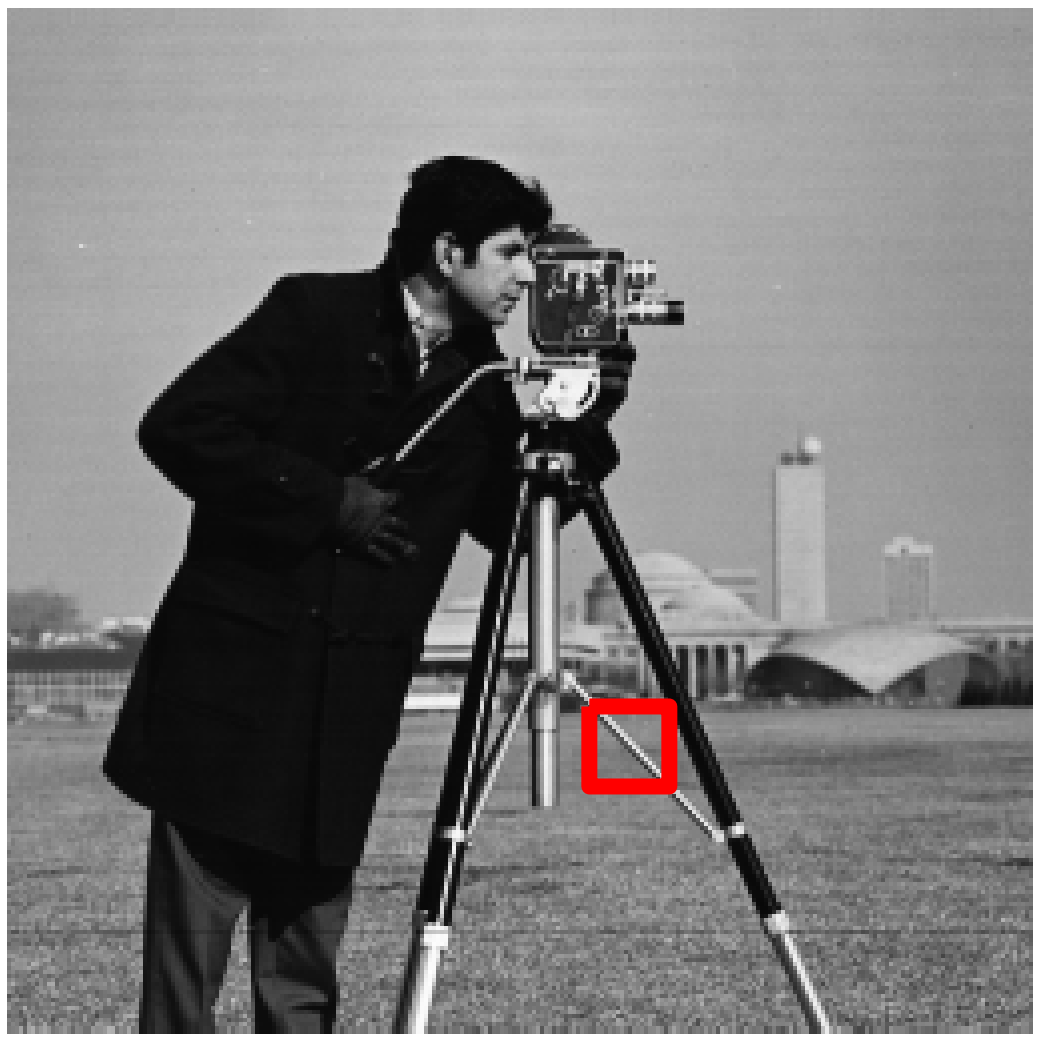

(a) Original low resolution image.

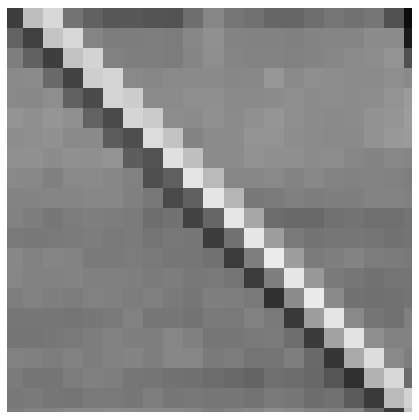

(b) Rectangular kernel.

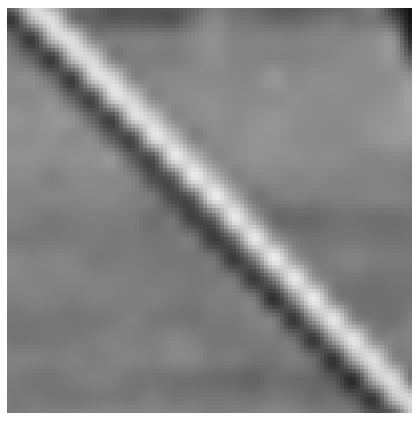

(c) Bicubic kernel.

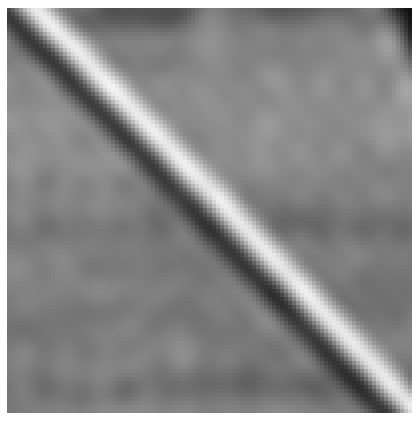

(d) Sinc kernel.

Fig. 14: Periodic structure in an interpolated signal is a phenomena related to the effective bandwidth of the interpolation kernel. The larger the portion of its energy outside $[-\pi, \pi]$, the stronger the periodic correlation. The three images on the right were obtained by scaling a patch of the original image by a factor of $5$ using three different methods. The portion of energy in the range $[-\pi, \pi]$ of the kernels is: (b) rectangular kernel - 61%; (c) bicubic kernel - 91%; (d) sinc - 100%. Suppressed periodic correlation, however, does not necessarily imply that the reconstruction error is small.

[48]:

$$\text{MSE}_A = E\left[\int_{t_0}^{t_0+1} |x(t) - \hat{x}(t)|^2 dt\right]. \tag{43}$$

It turns out that minimization of the average MSE leads to a correction filter independent of $t_0$ [48]. The second approach makes use of the fact that the best possible approximation to $x(t)$ in $\mathcal{W}$ is $P_{\mathcal{W}}x(t)$. Therefore this method aims at minimizing the projected MSE, defined as the MSE with respect to the optimal approximation in $\mathcal{W}$ [50]:

$$\text{MSE}_P = E\left[|P_{\mathcal{W}}x(t) - \hat{x}(t)|^2\right]. \tag{44}$$

Interestingly, both error measures (43) and (44) lead to the same digital correction filter, which is given by [48], [50]

$$H(e^{j\omega}) = \frac{\phi_{W\tilde{W}}\left(e^{j\omega}\right)}{\phi_{S\tilde{W}}\left(e^{j\omega}\right)\phi_{WW}\left(e^{j\omega}\right)}, \tag{45}$$

where here $\tilde{W}(\omega) = S(\omega)\Lambda_{xx}(\omega)$. This is also the solution obtained by the minimax regret criterion (see (37)) where $|L(\omega)|^{-2}$ replaces the spectrum $\Lambda_{xx}(\omega)$. Therefore, here again, $L(\omega)$ plays the role of the whitening filter of $x(t)$.

The average MSE criterion (43) can also be used to handle the dense grid recovery setup of Fig. 4. Minimization of the average MSE, in this case, leads to the corresponding minimax regret solution (39) with $\tilde{W}(\omega) = S(\omega)\Lambda_{xx}(\omega)$.

The mathematical equivalence between the minimax and Wiener formulas suggests selecting an "optimal" operator $L(\omega)$ in the minimax formulation that "whitens" the signal. In practice, one can either choose $L(\omega)$ in advance to approximately whiten signals typically encountered in a specific application (*e.g.*, magnetic resonance images), or specify a parametric form for $L(\omega)$ and optimize the parameters based on the samples $c[n]$ [32].

## V. Sparsity Priors

Before concluding this review, we briefly address sampling of sparse analog signals, a topic that has gained considerable interest in recent years.

Throughout the review we focused mainly on linear interpolation techniques which were sufficient to recover, or properly approximate, many classes of signals. An important case where nonlinear methods are necessary is when the prior on the signal is not a subspace but rather a sparsity constraint. For example, we may know that the signal has the form

$$x(t) = \sum_{k=1}^{N-1} a_k g(t - t_k) \tag{46}$$

for some coefficients $a_k$ and time instants $t_k$. Such a signal is said to have a sparse representation since only a few parameters are needed to specify it [22], [23]. If the values $t_k$ are known then this is just a subspace prior. More interesting is when the times $t_k$ need to be estimated along with the coefficients $a_k$. Several sampling strategies to deal with these signal classes have been suggested, known as finite rate of innovation sampling. It turns out that roughly $2N$ samples are enough to recover the entire signal with proper post processing.

Another important class of sparse signals is the class of signals whose frequency transform (or any other transform) has a multiband structure. In this case the Fourier transform consists of a finite number $N$ of bands, each of width at most $B$, as illustrated in Fig. 15. If the band locations $a_i, b_i$ are known, then this corresponds to a subspace prior and the methods discussed in this review can be used to recover the signal from its samples [59], [60], [61]. A very interesting question is whether the signal can be recovered at a rate lower than the Nyquist rate, $1/T$ in the figure, when the band locations are unknown. Such a sampling scheme is referred to as blind since it does not require knowledge of the band locations in the sampling and reconstruction stages.

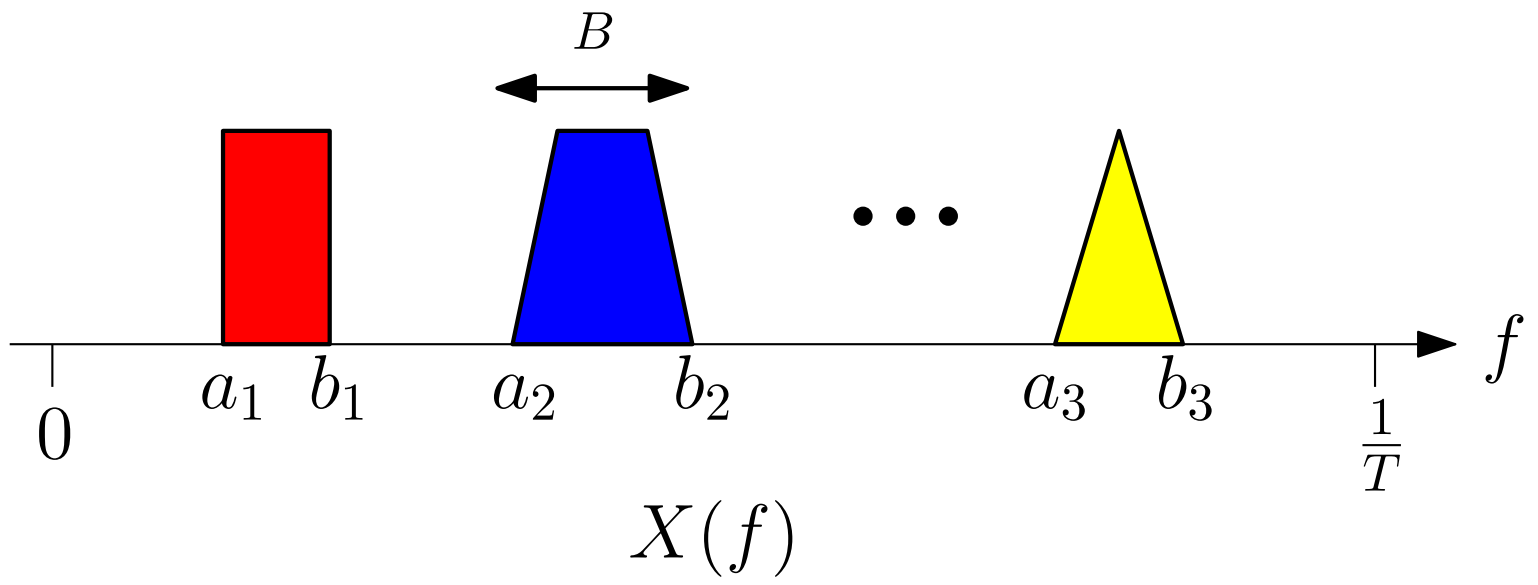

Fig. 15: Typical spectrum support of a multiband signal.

At first, it may seem that since the band locations are unknown, the signal can have energy in the entire Nyquist frequency range, and therefore lower than Nyquist sampling will not be sufficient to recover the signal. Surprisingly, this reasoning is in fact incorrect. In practice such classes of signals can be sampled at rates much lower than Nyquist, without impairing the ability to perfectly recover the signals [18]. The tradeoff is in that the reconstruction involves nonlinear processing of the samples. In fact, it can be proven that the minimal sampling rate at which such signals can be processed is twice the sum of the widths of the frequency bands, which can be significantly smaller than twice the highest frequency, corresponding to the Nyquist rate. When the band locations are known, the minimal sampling rate is exactly the sum of the bands, referred to as the Landau rate. Thus, the price for constructing an ideal blind system is only a factor of two. (In the presence of noise and other distortions, a larger factor will be necessary to ensure stable recovery.)

The techniques developed to sample and reconstruct such classes of signals are based on ideas and results from the emerging field of compressed sensing [16], [17]. However, while the latter deals with sampling of finite vectors, the multiband problem is concerned with analog sampling. By using several tools, developed in more detail in [19], [20], [21], it is possible to extend the essential ideas of compressed sensing to the analog domain. In this setting, the measurement matrix of traditional compressed sensing is replaced by a bank of analog filters. This broader framework combines ideas presented in this review with traditional compressed sensing tools in order to treat more general classes of signals, such as signals that lie in a union of SI spaces. In this case,

$$x(t) = \sum_{k=1}^{K} \sum_{n=-\infty}^{\infty} d_k[n] a_k(t-n), \tag{47}$$

for a set of generators $a_k(t)$ where only $M < K$ out of the sequences $d[n]$ are not identically zero. This model subsumes the bandlimited class of functions as a special case. These results can also be applied more generally to signals that lie in a union of subspaces [62], [63], which are not necessarily shift invariant.

## VI. Unified View and Practical Considerations

In this article, we reviewed a panorama of methods for recovering a signal from its samples. The solutions emerged from different assumptions on the underlying signal, the sampling process and the reconstruction mechanism. Whereas it is generally well understood how to model the sampling process in real-world applications, choosing the signal prior and the reconstruction scheme is typically left to the practitioner. These components affect both the performance and the computational load of the resulting algorithm.

Below, we emphasize commonalities and equivalence between the different methods in order to help the practitioner design the most appropriate filter for a particular application. We focus on the linear sampling scheme of Fig. 1 and on the reconstruction method of Fig. 3, in which the sampling and interpolation grids coincide. Similar considerations can be applied in the more general scenarios we surveyed as well.

The linear recovery algorithms corresponding to the first two columns of Table III share a common structure.

TABLE IV: Prior filter for different setups

| | **Subspace Prior** | **Smoothness Prior** | **Stochastic Prior** |
|---|---|---|---|
| **Assumption** | $x(t) = \sum d[n]a(t-n)$ | $\int \|L(\omega)X(\omega)\|^2 d\omega \leq U$ | $x(t)$ WSS with PSD $\Lambda_{xx}(\omega)$ |
| **Prior Filter** | $A(\omega)$ | $S(\omega)/\|L(\omega)\|^2$ | $S(\omega)\Lambda_{xx}(\omega)$ |

The digital correction filter $H(e^{j\omega})$ of Fig. 3 can be written in all cases in the form

$$H\left(e^{j\omega}\right) = \frac{\phi_{WP}\left(e^{j\omega}\right)}{\phi_{SP}\left(e^{j\omega}\right)\phi_{WW}\left(e^{j\omega}\right)}, \tag{48}$$

where $\phi_{SP}(e^{j\omega})$ is defined by (10). Here $S(\omega)$ and $W(\omega)$ are the CTFTs of the sampling and reconstruction filters, and $P(\omega)$, referred to as the prior filter, shapes the spectrum of $\hat{x}(t)$ according to the prior. The different priors together with the corresponding filters are summarized in Table IV.

The reconstruction filter $W(\omega)$ can either be chosen in advance to lead to efficient implementation, or optimized according to the prior. The solutions in the unconstrained setting can be recovered from (48) with $W(\omega) = P(\omega)$. Indeed, in this case, the filter of (48) reduces to

$$H\left(e^{j\omega}\right) = \frac{1}{\phi_{SP}\left(e^{j\omega}\right)}. \tag{49}$$

Substituting the values of the prior filter $P(e^{j\omega})$ into (49) according to Table IV leads to the first column of Table III. This filter also guarantees perfect reconstruction for any signal lying in the SI space spanned by the functions $\{p(t-n)\}$, offering an additional viewpoint on the prior filter $P(\omega)$: It defines the SI space for which perfect reconstruction is obtained using (49).

When the reconstruction filter $W(\omega)$ is fixed in advance, substituting the values of $P(\omega)$ from Table IV into (48) results in the second column of Table III, with the minimax solution in the case of a smoothness prior. The consistent solution follows from choosing $P(\omega) = W(\omega)$.

In general, practical evaluation of (48) may be difficult. One brute-force technique is to truncate the infinite series in (10) required for the computation of $\phi_{SP}(e^{j\omega})$, $\phi_{WP}(e^{j\omega})$ and $\phi_{WW}(e^{j\omega})$. Any filter design technique can then be used to approximate this desired response with an FIR or IIR filter. There are however important cases in which $H(e^{j\omega})$ can be determined analytically. An important example is when $s(t)$, $w(t)$ and $p(t)$ are all B-splines [64], [12]. The numerator in (48) then corresponds to an FIR filter with a simple formula. Each of the terms in the denominator correspond to a concatenation of a causal and anti-causal IIR filter. We may therefore first filter the data with a recursive formula running from left to right and then filter the result with the same formula running from right to left [64].

The unified interpretation of the different interpolation algorithm highlights the fact that choosing a specific method is equivalent to choosing the prior filter. This filter, in turn, should be matched to the typical frequency content of the input signals. In addition, we have also seen that a general purpose recovery algorithm (*i.e.*, one which can handle resampling at arbitrary points) requires an explicit expression for $w(t)$ in the time domain. This should be taken into consideration when choosing the prior in the unconstrained approach since $w(t) = p(t)$ in this case. Therefore, a kernel $p(t)$ with an analytic formula is beneficial.

Various priors have been previously proposed in the literature. A common choice corresponds to selecting $L(\omega) = a_0 + a_1 j\omega + a_2(j\omega)^2 + \cdots + a_K(j\omega)^K$ as a differential operator together with $S(\omega)$ chosen as a rational causal filter [50]. Another type of prior is associated with a family of WSS processes including the Matérn class [33]. The regularization filter in this case is given by $L(\boldsymbol{\omega}) = \prod_{m=1}^{K}(a_m + \|\boldsymbol{\omega}\|^2)^{\gamma_m}$, where $\boldsymbol{\omega}$ denotes the 2D frequency vector, $a_m > 0$ and $\gamma_m > 1$. The resulting kernel $p(\boldsymbol{t})$ was shown to have a closed form in the case of pointwise sampling (*i.e.*, $S(\boldsymbol{\omega}) = 1$) [65]. A non-isotropic version of this kernel was also used in [32], where the authors optimized the model parameters based on the samples $c[\boldsymbol{n}]$.

Finally, we comment briefly on the reconstruction filter $w(t)$. The key consideration in choosing $w(t)$ is its support, which determines the number of coefficients of the corrected sequence $d[n]$ participating in computing $\hat{x}(t_0) = \sum d[n]w(t_0 - n)$. Typically, kernels with support up to $4$ are used, requiring $4$ multiplications per time instant $t_0$ to compute $\hat{x}(t_0)$ (or $16$ in two dimensions). These include B-splines of degree $0$ to $3$ whose supports are $1$ to $4$ respectively, the Keys cubic interpolation kernel [66] whose support is $4$, and the Lanczos kernel with support $4$. Some of the commonly used kernels, such as Keys and Lanczos, possess the interpolation property, namely $w(n) = \delta[n]$. This implies that if we are given pointwise samples of the signal $c[n] = x(n)$, then no correction filter $h[n]$ is needed in order to obtain a consistent reconstruction satisfying $\hat{x}(n) = c[n]$.

## Appendix A

## Box A: Basis Expansions

A Schauder basis for a complex Hilbert space $\mathcal{H}$ is a countable set of vectors $\{x_n\}$ in $\mathcal{H}$ such that every vector $x \in \mathcal{H}$ can be written uniquely as a series

$$x = \sum_{n=-\infty}^{\infty} c[n]x_n \tag{50}$$

with scalars $c[n]$. For example, the set of complex exponentials $x_n(t) = \exp\{j\omega nt\}$ defined over $t \in [-\pi, \pi]$ is a Schauder basis for the space $L_2[-\pi, \pi]$ of square integrable functions over $[-\pi, \pi]$. In this basis, the expansion coefficients $c[n]$ of a function $x(t)$ are its Fourier coefficients.

A countable set of vectors $\{x_n\}$ in $\mathcal{H}$ is a Riesz basis for $\mathcal{H}$ if it is complete and there exist two constants $\alpha > 0$ and $\beta < \infty$ such that

$$\alpha \sum_{n=-\infty}^{\infty} |c[n]|^2 \leq \left\| \sum_{n=-\infty}^{\infty} c[n]x_n \right\|^2 \leq \beta \sum_{n=-\infty}^{\infty} |c[n]|^2, \quad \forall c \in \ell_2. \tag{51}$$

Here $\|y\|$ is the norm over $\mathcal{H}$. Riesz bases have the desired stability property, namely that a slight change in the expansion coefficients $c[n]$ is ensured to entail only a small change in $x$. Consequently, these bases are important in ensuring stable sampling schemes.

An important question is how to obtain the expansion coefficients $c[n]$ of a vector $x$. If the basis $\{x_n\}$ is orthonormal (*i.e.*, $\langle x_m, x_n \rangle = \delta_{mn}$ where $\langle x, y \rangle$ is the inner product over $\mathcal{H}$) then $c[n] = \langle x, x_n \rangle$. This follows from taking the inner products of both sides of (50) with $x_m$ and exploiting the orthogonality property. To determine the expansion coefficients when using a general non-orthogonal basis, we follow a similar route using the biorthogonal vectors, or dual basis $\tilde{x}_n$. The dual basis of $x_n$ is the unique basis of $\mathcal{H}$ that satisfies the property

$$\langle x_m, \tilde{x}_n \rangle = \delta_{mn}. \tag{52}$$

Taking the inner products of both sides of (50) with respect to $\tilde{x}_m$, we find that

$$c[n] = \langle x, \tilde{x}_n \rangle. \tag{53}$$

If $x_n$ is a Riesz basis, then so is its biorthogonal basis.

When the set of vectors $\{x_n\}$ span only a subspace $\mathcal{U}$ of $\mathcal{H}$, there may be many choices of biorthogonal bases in $\mathcal{H}$ satisfying (52). Intuitively, the biorthogonal basis vectors should span a subspace with the same number of degrees of freedom as $\mathcal{U}$. A formal statement of this observation is that given any subspace $\mathcal{V}$ satisfying the direct sum condition $\mathcal{H} = \mathcal{U} \oplus \mathcal{V}^{\perp}$, there exists a unique set of vectors $\{\tilde{x}_n\}$ lying in $\mathcal{V}$ which constitute a biorthogonal basis for $\{x_n\}$. This set is called the oblique dual basis of $\{x_n\}$ in $\mathcal{V}$ [14], [15], [13], [42], [44], [31]. The vectors $\{\tilde{x}_n\}$ satisfy (52) and form a basis for $\mathcal{V}$, that obey the Riesz condition given that $\{x_n\}$ is a Riesz basis. In each subspace $\mathcal{V}$ there is only one dual basis. The canonical dual basis refers to the choice $\mathcal{U} = \mathcal{V}$. This concept can also be extended to sets of vectors that are linearly dependent, leading to oblique dual frames.

## Appendix B

## Box B: Discrete and Continuous Fourier Transforms

The continuous-time Fourier transform (CTFT) of a signal $x(t)$ in $L_2$ is defined as

$$X(\omega) = \int_{-\infty}^{\infty} x(t) e^{-j\omega t} dt. \tag{54}$$

We use the convention that upper case letters denote Fourier transforms. The discrete-time Fourier transform (DTFT) of a sequence $x[n]$ in $\ell_2$ is defined by

$$X(e^{j\omega}) = \sum_{n=-\infty}^{\infty} x[n] e^{-j\omega n}. \tag{55}$$

The DTFT is $2\pi$-periodic; to emphasize this fact we use the notation $X(e^{j\omega})$.

The DTFT of the sampled sequence $y(t=n)$ is related to the CTFT of $y(t)$ by

$$Y(e^{j\omega}) = \sum_{k=-\infty}^{\infty} Y(\omega - 2\pi k). \tag{56}$$

In the reverse direction, if the sequence $d[n]$ is used to create a continuous-time signal $f(t) = \sum_n d[n] y(t-n)$, then

$$F(\omega) = D(e^{j\omega}) Y(\omega). \tag{57}$$

An important sequence encountered in signal recovery problems is the sampled cross correlation $r_{as}[n] = \langle a(t), s(t-n) \rangle$. This sequence can be obtained by sampling the output of the filter $s(-t)$ with $a(t)$ as its input.

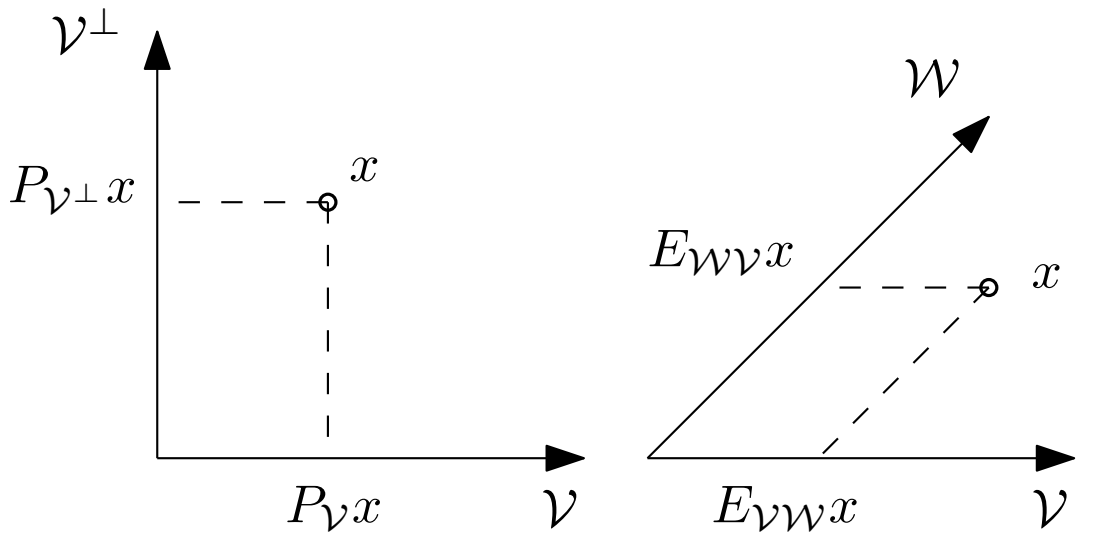

(a) Orthogonal projection. (b) Oblique projection.

Fig. 16: Decomposition of a vector $x$ into two components using an orthogonal projection and an oblique projection.

An immediate consequence from (56) is that the DTFT of $r_{as}[n]$ can be expressed as

$$\phi_{SA}(e^{j\omega}) = \sum_{k=-\infty}^{\infty} S^*(\omega - 2\pi k)A(\omega - 2\pi k), \tag{58}$$

where $(\cdot)^*$ denotes the complex conjugate.

The set $\{a(t-n)\}$ is orthonormal if each function $a(t-n)$ is orthonormal to all of its integer shifts. This is equivalent to requiring that $r_{aa}[n] = \delta[n]$ where

$$\delta[n] = \begin{cases} 1 & n = 0; \\ 0 & \text{otherwise}. \end{cases} \tag{59}$$

From (58) we conclude that $\{a(t-n\}$ is an orthonormal sequence if and only if

$$\phi_{aa}(e^{j\omega}) = \sum_{k=-\infty}^{\infty} |A(\omega - 2\pi k)|^2 = 1. \tag{60}$$

## Appendix C

## Box C: Projections in Hilbert Spaces

A projection $E$ in a Hilbert space $\mathcal{H}$ is a linear operator from $\mathcal{H}$ onto itself that satisfies the property

$$E^2 = E. \tag{61}$$

The importance of projection operators is that they map the entire space $\mathcal{H}$ onto the range space $\mathcal{R}(E)$, and leave vectors in this subspace unchanged. Furthermore, property (61) implies that every vector in $\mathcal{H}$ can be uniquely written as the combination of a vector in $\mathcal{R}(E)$ and a vector in the null space $\mathcal{N}(E)$, that is, we have the direct sum decomposition $\mathcal{H} = \mathcal{R}(E) \oplus \mathcal{N}(E)$. This is illustrated in Fig. 16 for two different projection operators. Therefore, a projection is completely determined by its range space and null space.

An orthogonal projection $P$ is a Hermitian projection operator. In this case the range space $\mathcal{R}(P)$ and null space $\mathcal{N}(P)$ are orthogonal. Therefore, an orthogonal projection is completely determined by its range space. We use the notation $P_{\mathcal{V}}$ to denote an orthogonal projection with range $\mathcal{V} = \mathcal{R}(P_{\mathcal{V}})$.

An oblique projection $E_{\mathcal{V}\mathcal{W}}$ is an operator satisfying the projection property (61) that is not Hermitian. Its range space is given by $\mathcal{V}$ so that $E_{\mathcal{V}\mathcal{W}}x = x$ for any $x \in \mathcal{V}$, and its null space is given by $\mathcal{W}$ so that $E_{\mathcal{V}\mathcal{W}}x = 0$ for any $x \in \mathcal{W}$.

When decomposing the space using an orthogonal projection, the vectors comprising the decomposition are orthogonal, since $\mathcal{R}(P)$ and $\mathcal{N}(P)$ are orthogonal spaces. This is not true when using an oblique projection, as illustrated in Fig. 16. Another important feature of the orthogonal projection is that the norm of the projection is never larger than the original norm:

$$\|P_{\mathcal{V}}x\| \leq \|x\|. \tag{62}$$

This inequality does not necessarily hold for an oblique projection. In fact, the norm of the oblique projection can be much larger than the signal norm. Consequently, algorithms relying on the oblique projection can cause a significant increase in the noise if it is not constrained to the range space of the projection. On the other hand, orthogonal projections are more stable in the presence of noise due to (62).